# Landau-Quantized Excitonic Absorption and Luminescence in a Monolayer Valley Semiconductor

Erfu Liu[1], Jeremiah van Baren[1], Takashi Taniguchi[2], Kenji Watanabe[2], Yia-Chung Chang[3], Chun Hung Lui[1]*

[1] Department of Physics and Astronomy, University of California, Riverside, CA 92521, USA.
[2] National Institute for Materials Science, Tsukuba, Ibaraki 305-004, Japan
[3] Research Center for Applied Sciences, Academia Sinica, Taipei 11529, Taiwan
* Corresponding author. Email: joshua.lui@ucr.edu

**Abstract:**

We investigate Landau-quantized excitonic absorption and luminescence of monolayer $WSe_2$ under magnetic field. We observe gate-dependent quantum oscillations in the bright exciton and trions (or exciton-polarons) as well as the dark trions and their phonon replicas. Our results reveal spin- and valley-polarized Landau levels (LLs) with filling factors $n = +0, +1$ in the bottom conduction band and $n = -0$ to $-6$ in the top valence band, including the Berry-curvature-induced $n = \pm 0$ LLs of massive Dirac fermions. The LL filling produces periodic plateaus in the exciton energy shift accompanied by sharp oscillations in the exciton absorption width and magnitude. This peculiar exciton behavior can be simulated by semi-empirical calculations. The experimentally deduced g-factors of the conduction band ($g \sim 2.5$) and valence band ($g \sim 15$) exceed those predicted in a single-particle model ($g = 1.5, 5.5$, respectively). Such g-factor enhancement implies strong many-body interactions in gated monolayer $WSe_2$. The complex interplay between Landau quantization, excitonic effects, and many-body interactions makes monolayer $WSe_2$ a promising platform to explore novel correlated quantum phenomena.

**Main text:**

Monolayer transition metal dichalcogenides (TMDs, e.g. $MoS_2$ and $WSe_2$) host tightly bound excitons in two time-reversal valleys (K, K') with opposite spins, magnetic moment, Berry curvature, and circular dichroism [1-14]. With the application of magnetic field, the two valleys exhibit opposite Zeeman shift [15-19] and contrasting Landau levels (LLs) [20-22], including two LLs with filling factor $n = 0$ due to the Berry curvature of massive Dirac fermions [22-24] [Fig. 1(a)]. These two LLs are most relevant to the fractional quantum Hall effect and Wigner crystallization, which should first occur near the lowest LLs [25-28]. Besides, low-lying LLs contribute to the exciton formation. Given the small LL spacing (~10 meV) in TMDs due to the large carrier mass, the exciton with binding energy > 150 meV is the superposition of many LLs. These distinctive features make monolayer TMDs an excellent system to explore novel exciton-LL coupling and many-body interactions in two dimensions.

Much interesting LL physics in TMDs has been revealed by transport, scanning probe, optical and single-electron-transistor spectroscopy [29-39]. But these experiments are subject to different limitations. For instance, the high contact resistance and poor TMD material quality have limited transport probes to LLs with $n \geq 3$ in few-layer TMDs and $n \geq 17$ in monolayer TMDs [29-37]. Non-optical probes cannot assess the excitonic states. Recent optical experiments reported inter-LL optical transitions in monolayer WSe$_2$, but only in the high-charge-density regime where the excitonic effect is much reduced [38]. T. Smolenski *et al* recently reported excitonic quantum oscillations in monolayer MoSe$_2$, but only for absorption and $n \geq 1$ LLs of bright excitonic states [40]. More comprehensive studies are needed to explore the Landau-quantized excitonic physics for bright and dark excitonic states, with optical absorption and emission, on the electron and hole sides, and in a broad LL spectrum including the $n = 0$ LLs.

In this Letter, we investigate Landau-quantized excitonic absorption and luminescence in monolayer WSe$_2$ under magnetic fields up to B = 17.5 T. We observe gate-dependent quantum oscillations in the bright exciton and trions as well as the dark trions and their phonon replicas (here trions refer broadly to correlated states between excitons and Fermi sea, including exciton-polarons [41-43]). Our results reveal spin- and valley-polarized LLs with $n = –6$ to $+1$ from the hole to electron sides, including the two valley-contrasting $n = 0$ LLs [Fig. 1(b)]. The gate-dependent exciton energy shift exhibits periodic plateaus, accompanied by sharp oscillations in the exciton absorption width and magnitude. Such unusual exciton behavior can be simulated by semi-empirical calculations. The g-factors of the conduction and valence bands (g ~ 2.5, 15 respectively) exceed the predicted values in a single-particle model (g = 1.5, 5.5 respectively), implying strong many-body interactions in the system. Our research reveals complex interplay between Landau quantization, excitonic effects, and many-body interactions in monolayer WSe$_2$.

We investigate ultraclean monolayer WSe$_2$ gating devices encapsulated by hexagonal boron nitride on Si/SiO$_2$ substrates [44-47]. All the experiments were conducted at temperature T ~ 4 K. To study the absorption properties, we measure the reflectance contrast $\Delta R/R = (R_s – R_r)/R_r$, where $R_s$ is the reflection spectrum on the sample position and $R_r$ is the reference reflection spectrum on a nearby area without WSe$_2$. We selectively probe the optical transitions in the K and K' valleys with right- and left-handed circular polarization, respectively [11-14]. At zero magnetic field, the two valleys, being energy-degenerate, exhibit the same gate-dependent $\Delta R/R$ map featuring the A-exciton ($A^0$) and A-trions ($A^-$, $A^+$) [Fig. 2(a)]. At out-of-plane magnetic fields up to B = 17.5 T, however, the K and K' valleys exhibit distinct $\Delta R/R$ response. While the K'-valley map is qualitatively similar to that at zero magnetic field [Fig. 2(b)], the K-valley map exhibits several differences [Fig. 2(c-d)]. First, the $A^+$ trion is quenched. This is because the K valley lies higher than the K' valley under the magnetic field; the injected holes only fill the K valley [Fig. 1(a)]. The Pauli exclusion principle forbids the formation of $A^+$ trion with two holes in the K valley. Second, with the suppression of $A^+$ trion, the K-valley $A^0$ exciton is retained on the hole side, where it exhibits pronounced oscillations with increasing hole density. Third, the $A^-$ trion $\Delta R/R$ response also oscillates as the electron density increases. Fourth, the period of these oscillations increases linearly with the magnetic field [Fig. 2(c-e)]. These results mimic the Shubnikov-de Hass oscillations in quantum transport and strongly suggest the formation of LLs [29-39].

From the reflectance contrast maps, we extract the real part of the optical sheet conductivity (σ) of monolayer WSe$_2$ by solving the optical problem in our device structure (see Supplemental Material [47]). σ is proportional to the optical absorption of monolayer WSe$_2$. Fig. 3(a) displays the σ map of the K-valley $A^0$ exciton at B = 17.5 T. Fig. 3(b-d) show the extracted exciton peak energy, peak width, peak conductivity and integrated conductivity as a function of gate voltage ($V_g$). Remarkably, periodic plateaus appear as the exciton energy blueshifts on the hole side. At the step edges, the exciton peak is maximally broadened, and the peak conductivity drops to a local minimum. But the integrated conductivity drops relatively smoothly with $V_g$.

These quantized and oscillatory excitonic features can be qualitatively understood by the LL formation. At zero magnetic field, the hole injection gradually increases the screening and state-filling effects. This will renormalize the band gap and reduce the exciton binding energy, leading to a smooth blueshift of exciton energy [42, 43, 77, 78]. The smooth process will be disrupted when LLs are formed. When the Fermi energy ($E_F$) varies within a LL gap, the screening and state-filling effects remain unchanged; the exciton energy will not shift with $V_g$ and hence a plateau appears; the LL gap will also suppress the carrier scattering and narrow the exciton peak. But when the Fermi energy varies within a LL with singular density of states, the rapid increase of hole density will drastically increase the screening and state-filling effects, leading to a step-like exciton blueshift. The phase space in a partially filled LL will also facilitate the carrier scattering and hence broaden the exciton peak and reduce the peak conductivity value. Based on this analysis, we use the exciton width maxima and conductivity minima to identify the $V_g$ positions of half-filled LLs [Fig. 2(c-d)].

The above qualitative picture can be quantified by semi-empirical calculations. We have calculated the exciton conductivity spectrum by solving the massive Dirac equation for the electron and hole, including their screened Coulomb interaction, for monolayer $WSe_2$ under magnetic field (see Supplemental Material [47]). We only consider the hole filling in the K valley. Our calculations reproduce the main observation and confirm the assignment of half-filled LL positions [Fig. 3(e-h)].

We next discuss the reflection results on the electron side. The electrons are injected into the K' valley, which lies below the K valley, in the conduction band under the magnetic field [Fig. 1] [15]. The K'-valley electrons have no state-filling effect on the K-valley excitons, though they can screen the excitonic interaction. In our results, the K-valley $A^0$ exciton shows no oscillation on the electron side [Fig. 2(d)]. This indicates weak influence of the K'-valley LL filling on the K-valley excitons.

The LL filling in the K' conduction valley, however, is found to induce noticeable quantum oscillations in the $A^-$ intervalley trion, which couples a K-valley exciton and the K'-valley Fermi sea. The $A^-$ trion exhibits two kinks in the conductivity and resonance energy when $V_g$ increases [Fig. 4]. We assign these kinks as the half-filled LL positions. When the Fermi energy lies within a LL, the phase space and free carriers in the partially filled LL can facilitate the trion formation, enhance the trion oscillator strength, and renormalize the trion energy. When the Fermi energy lies within a LL gap, the suppression of carrier scattering can hinder the trion formation.

Our observed exciton and trion oscillations allow us to identify the $V_g$ positions of half-filled LLs in the K valence valley and lower K' conduction valley, respectively. Fig. 2(e) displays these LL $V_g$ positions as a function of magnetic field (see Supplemental Material for data at other magnetic fields [47]). By calculating the Landau fan diagram, we fit the data on both the electron and hole sides by using a BN dielectric constant 3.06 [57, 79] and the LL density of states with no spin and valley degeneracy [lines in Fig. 2(e)]. Comparison with theory confirms our observation of valence LLs with filling factors $n = –0$ to $–6$ and conduction LLs with $n = +0, +1$; all the observed LLs are spin- and valley-polarized. Notably, we directly resolve the two $n = 0$ LLs in the K valence valley and lower K' conduction valley [Fig. 1]. These two LLs are induced by the Berry curvature of massive Dirac fermions [22-24]. Evidence of the $n = 0$ LLs was previously given by inter-LL optical transitions [38]; but here we directly observe and gate-control the two $n = 0$ LLs.

Our discussion above is limited to absorption properties. We have also observed quantum oscillations in the excitonic photoluminescence (PL) [Fig. 5]. Fig. 5(a) displays the right-handed PL map at B = 17.5 T, which reveals the K-valley emission of the bright exciton ($A^0$) and trions ($A^-$, $A^+$). The $A^+$ trions emit weak but observable PL [Fig. 5(a)], unlike the complete $A^+$ suppression in absorption [Fig. 2(d)]. This is because the laser excitation can generate transient holes in the K' valley to form intervalley trions [Fig. 5(c)]. We observe oscillations in the PL intensity of both the $A^0$ exciton and $A^+$ trion from the K valley, but their oscillations are out of phase to each other [Fig. 5(b)]. The PL intensity of excitons and trions reflects their steady-state population in monolayer WSe$_2$. According to the exciton line width in Fig. 3(c), the exciton lifetime is significantly shortened at half LL filling due to the enhanced carrier scattering. This implies a higher exciton-to-trion relaxation rate at half LL filling. Moreover, the phase space in a partially filled LL favors the trion formation. We thus expect the exciton population to drop and the trion population to increase at half LL filling. This picture qualitatively explains the opposite oscillations of $A^0$ and $A^+$ PL intensity in our experiment. The $A^+$ PL maxima should denote the half-filled LL positions [Fig. 5(a-b)].

Fig. 5(d) displays the left-handed PL map at B = 17.5 T, which reveals the emission of bright excitons and trions from the K' valley [Fig. 5(f)]. On the hole side, the $A^0$ exciton is quenched and the $A^+$ trion dominates, because the injected holes in the K valley can efficiently couple to the K'-valley exciton to form intervalley trions [Fig. 5(f)]. The $A^+$ PL intensity oscillates as the hole density increases [Fig. 5(e)]. The $A^+$ PL maxima should correspond to partially filled LLs, whose free carriers and phase space favor the trion formation, like the $A^-$ oscillation in Fig. 4. Our experiment could not resolve any appreciable oscillation of $A^+$ PL energy and peak width.

Remarkably, our PL maps also reveal quantum oscillations of dark trions and their phonon replicas on the hole side [Fig. 5(a), (d); Fig. S13]. As the dark exciton ($D^0$) and trions ($D^+$, $D^-$) in monolayer WSe$_2$ emit linearly polarized light, our circularly-polarized PL maps show their emission from both valleys, which is split into two lines under magnetic field [45]. The lower-energy line ($D_l^+$) and all replicas on the hole side oscillate in phase with $A^+$ [Fig. 5, S14]. This is reasonable because partially filled LLs also favor the dark-trion formation, and the dark-trion replicas acquire oscillator strength from the $A^+$ state through trion-phonon interaction [46, 53]. But surprisingly, the higher-energy line ($D_h^+$) oscillates oppositely with $A^+$ and $D_l^+$ [Fig. S14]. We speculate that this is because the $D_l^+$ and $D_h^+$ states compete for the insufficient transient holes in the K' valley to form trions. When one prevails, the other is suppressed, thus leading to opposite oscillations between them (see more discussion in the Supplemental Material [47])

Our observation of numerous valley-polarized LLs reveals extraordinarily large valley Zeeman shift beyond a single-particle picture. Our reflectance contrast maps show seven LLs ($n = -0$ to -6) before the exciton is suppressed [Fig. 2(c-d)]. Our PL maps also show seven LLs ($n = -0$ to -6) before the $A^+$ PL is abruptly intensified and the $A^+$ oscillation is smoothened [Fig. S9, S11] [47]. Both indicate that the injected holes start to occupy the K' valley after the K valley. The energy mixing of two valleys can facilitate the carrier scattering, enhance the trion formation, and obscure the trion oscillation. Therefore, the K valence valley should be at least 6 LL spacings higher than the K' valence valley. The LL energy spacing is $2\mu_B B m_0/m^*$, where $\mu_B$ is the Bohr magneton, $m_0$ is the free electron mass, and $m^* \approx 0.4\, m_0$ is the hole effective mass in monolayer

WSe$_2$ [57]. The K valley is thus at least 30 $\mu_B B$ above the K' valley. The g-factor of the K valence valley is deduced to be ~15. This far exceeds the g-factor (5.5) predicted in a single-particle model [15, 47]. This deviation implies strong many-body interactions in the system.

Similar interaction-driven enhancement of g-factor is found in the conduction valleys. After filling the $n = +0, +1$ LLs in the K' valley, the $A^-$ feature is noticeably broadened and the quantum oscillation disappears [Fig. 4]. This signifies that the injected electrons start to occupy the K valley after the K' valley. The mixing of intra- and inter-valley trions broaden and smoothen the $A^-$ feature [Fig. S7]. We therefore deduce that the K conduction valley is at least 1 LL spacing higher than the K' conduction valley. Given the electron effective mass $m^* \approx 0.4 m_0$ [30, 57, 58], the K valley is at least $5\mu_B B$ above the K' valley. The g-factor of the K conduction valley is thus at least 2.5. This exceeds the g-factor (1.5) predicted by a single-particle model. Recent experiments also report g-factor enhancement in the conduction valleys of bilayer MoS$_2$ [37], monolayer and bilayer MoSe$_2$ [34, 80], and the valence valleys of monolayer WSe$_2$ [39, 81]. Here we further demonstrate that many-body interactions can enhance the g-factors in both the conduction and valence valleys of gated monolayer WSe$_2$ [Fig. 1(a-b)].

In summary, we have observed quantum oscillations in the absorption and luminescence of exciton and trion states in monolayer WSe$_2$ under magnetic field. The plateaus in the exciton energy shift indicate well-separated LLs. The enhanced g-factors imply strong many-body interactions. The combined observation of the lowest LLs and strong many-body interactions suggest that fractional quantum Hall states can emerge in monolayer WSe$_2$. In addition, the $n = 0$ half-filled LLs contain a very low density (~$1.4 \times 10^{11}$ cm$^{-2}$) of free carriers (not trapped by defects). This density corresponds to a large Wigner-Seitz radius $r_s \approx 25$ [47], close to the predicted condition $r_s \gtrsim 31$ of Wigner crystallization in two dimensions [62]. Therefore, our optical methods, inert to the contact resistance, may help probe the fractional quantum Hall states and Wigner crystals in monolayer valley semiconductors.

**Acknowledgement:** We thank Dmitry Smirnov and Zhengguang Lu for assistance in the experiments. A portion of this work was performed at the National High Magnetic Field Laboratory, which is supported by the National Science Foundation Cooperative Agreement No. DMR-1644779 and the State of Florida. Y.-C.C. is supported by Ministry of Science and Technology (Taiwan) under Grant Nos. MOST 107-2112-M-001-032 and MOST 108-2112-M-001-041. K.W. and T.T. acknowledge support from the Elemental Strategy Initiative conducted by the MEXT, Japan and the CREST (JPMJCR15F3), JST.

E. Liu and J. van Baren contributed equally to this work.

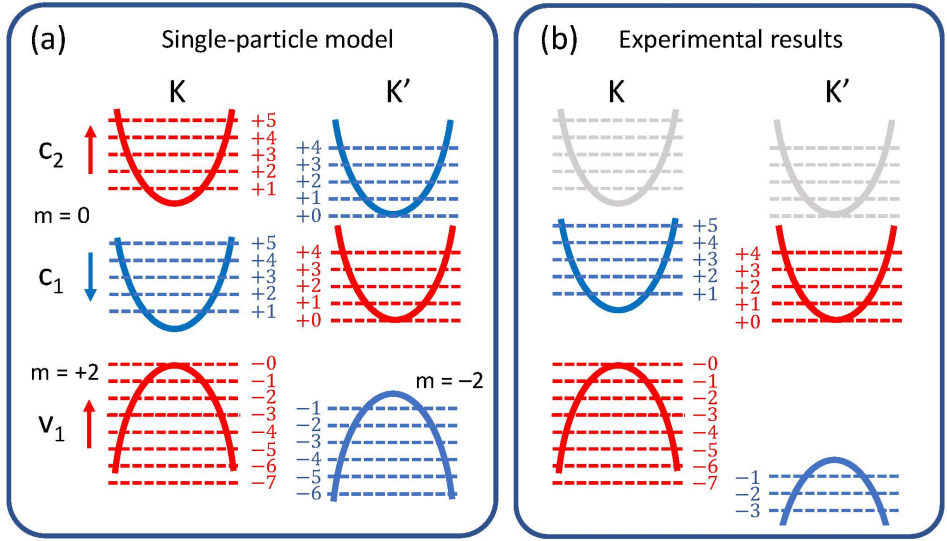

FIG. 1. (a) Schematic Landau levels (LLs, dashed lines) in monolayer WSe$_2$ predicted by a single-particle model. The arrows and color denote the electron spin in the conduction bands ($c_1$, $c_2$) and valence band ($v_1$). $m = 0, \pm 2$ are the azimuthal quantum numbers of the atomic orbits in the conduction bands and valence band, respectively. (b) LLs revealed by our experiment, indicating strong enhancement of valley Zeeman shift by many-body interactions.

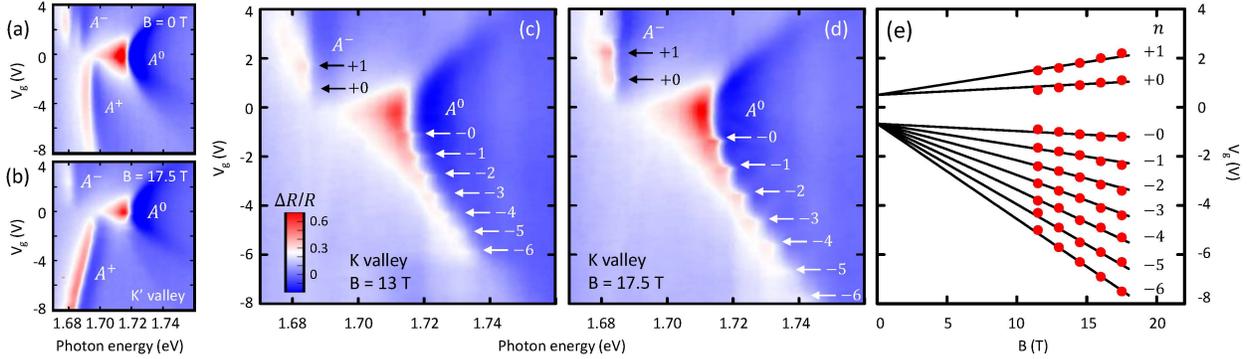

FIG. 2. (a) Gate-dependent $\Delta R/R$ reflectance contrast map of monolayer WSe$_2$ with no magnetic field. (b-d) Reflectance contrast maps in left-handed (b) and right-handed (c-d) circular polarization, corresponding to optical transitions in the K' and K valley, respectively, at magnetic fields B = 13.0 T (c) and B = 17.5 T (b, d). The arrows denote the quantum oscillations. (e) The measured gate voltages ($V_g$) at half-filled LLs as a function of magnetic field. The lines are the fitted Landau fan diagram. $n$ is the LL filling factors. Our fits show a $\Delta V_g \sim 1.2$ V gap between the electron and hole sides at B = 0 T, presumably due to the filling of defect states inside the band gap. The experiments were conducted at T ~ 4 K.

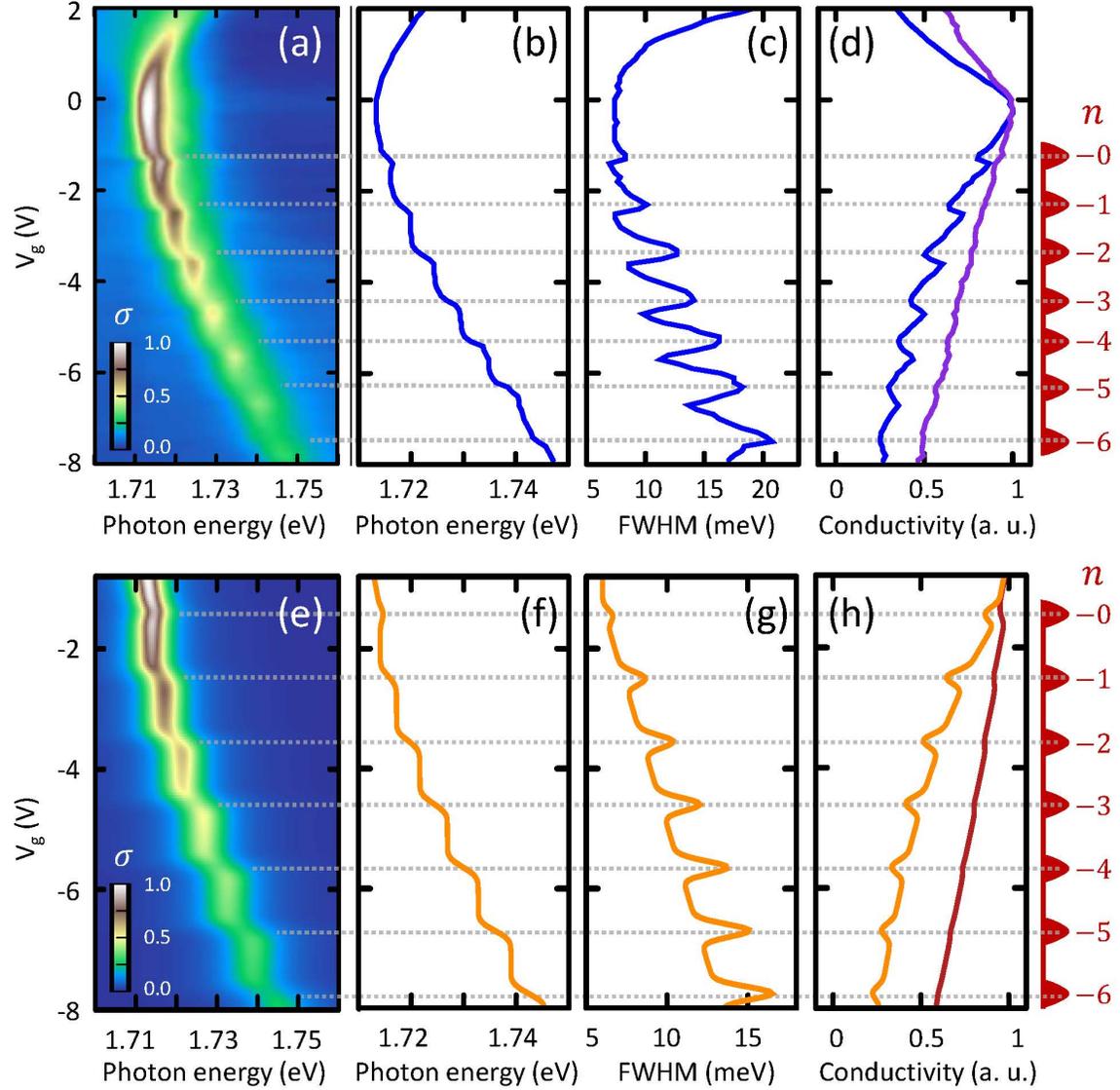

FIG. 3. (a) The optical sheet conductivity of $A^0$ exciton extracted from Fig. 2(d). (b) The exciton peak energy, (c) full width at half maximum (FWHM), (d) peak conductivity (blue) and integrated conductivity (purple), extracted from (a). (e) The theoretical exciton conductivity map, (f) peak energy, (g) FWHM, (h) peak conductivity (orange) and integrated conductivity (red). All conductivity values are normalized to one. The dashed lines denote the half-filled LL positions.

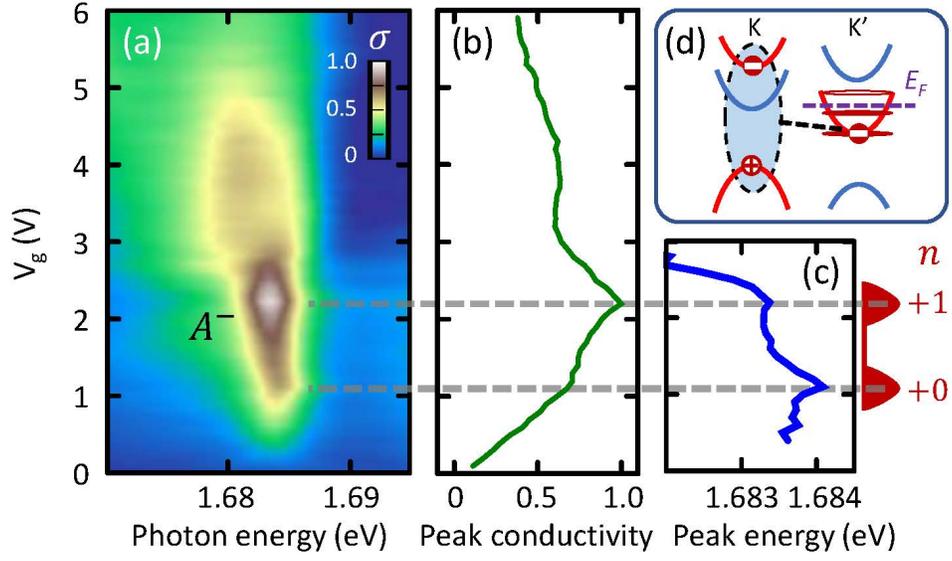

FIG. 4. (a) The normalized sheet conductivity of $A^-$ trions extracted from Fig. 2(d). (b) The trion peak conductivity and (c) peak energy extracted from (a). The dashed lines denote the half-filled LL positions. (d) The schematic band configuration of intervalley $A^-$ trion with the relevant LLs and Fermi energy ($E_F$).

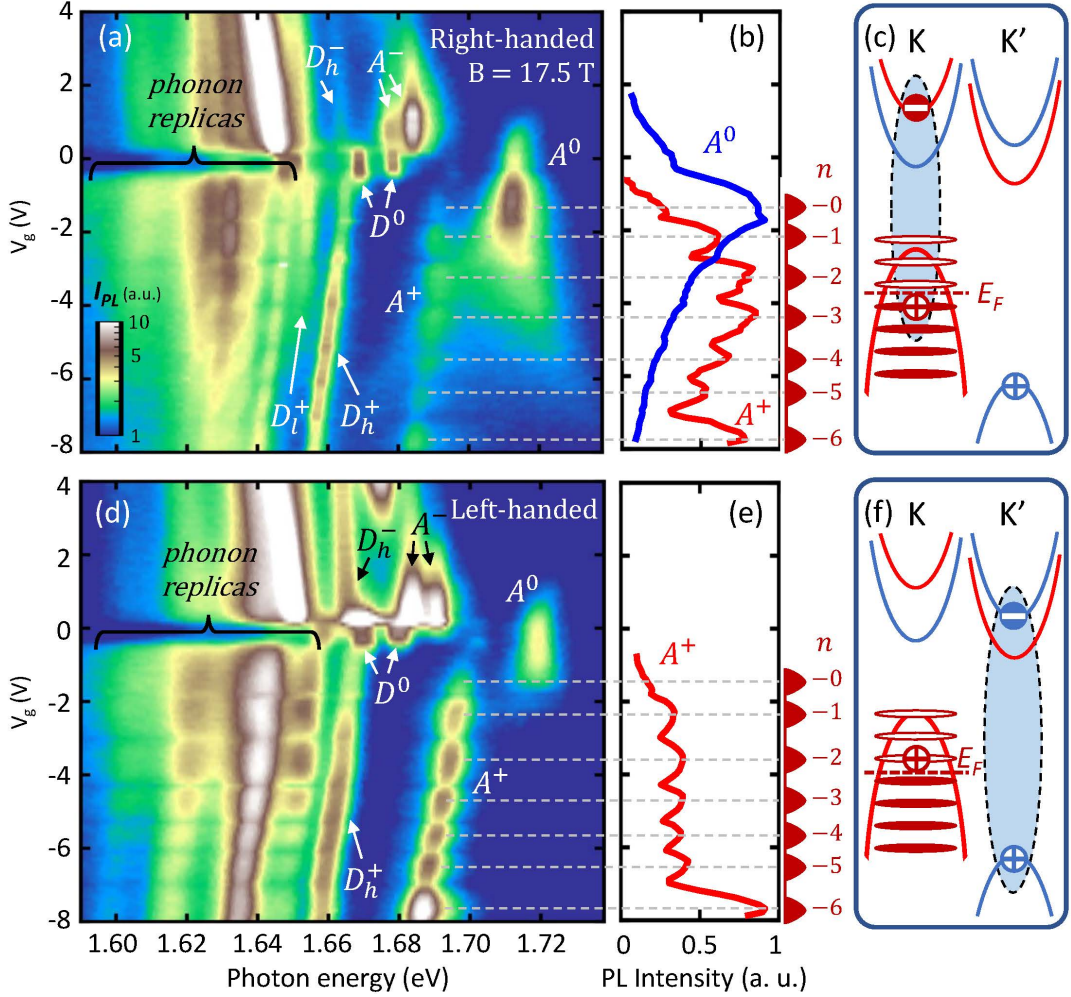

FIG. 5. (a) The gate-dependent photoluminescence (PL) map at right-handed circular polarization, which reveals the K-valley emission of bright excitonic states ($A$), under 532-nm continuous laser excitation and B = 17.5 T at 4K. (b) The normalized integrated PL intensity for the $A^0$ exciton (blue) and $A^+$ trion (red) in (a). The dashed lines denote the estimated half-filled LL positions. (c) The schematic band configuration of K-valley $A^+$ trion with the relevant LLs and Fermi energy ($E_F$) under the magnetic field. The K'-valley hole comes from the laser excitation. (d-f) Similar figures as (a-c) for left-handed circular polarization, which reveals the K'-valley emission of bright excitonic states. Panels (a) and (d) share the same color scale. (a) and (d) also show the emission of dark excitons ($D^0$) and dark trions ($D^+, D^-$) as well as a series of phonon replicas of the dark states. The circularly-polarized PL maps can reveal the emission of dark states from both valleys because their emission has linear polarization. Under the magnetic field that lifts the valley degeneracy, the dark trion emission $D^+$ ($D^-$) is split into a lower-energy line and a higher-energy line, which are denoted respectively as $D_l^+$ and $D_h^+$ on the hole side ($D_l^-$ and $D_h^-$ on the electron side).

# Supplemental Material for
# Landau-Quantized Excitonic Absorption and Luminescence in a Monolayer Valley Semiconductor

Erfu Liu[1], Jeremiah van Baren[1], Takashi Taniguchi[2], Kenji Watanabe[2], Yia-Chung Chang[3], Chun Hung Lui[1]*

[1] Department of Physics and Astronomy, University of California, Riverside, CA 92521, USA.
[2] National Institute for Materials Science, Tsukuba, Ibaraki 305-004, Japan
[3] Research Center for Applied Sciences, Academia Sinica, Taipei 11529, Taiwan
* Corresponding author. Email: joshua.lui@ucr.edu

**Table of contents:**

## 1. Device fabrication

We fabricate ultraclean monolayer WSe$_2$ devices encapsulated by hexagonal boron nitride (BN) using a standard micro-mechanical polymer-stamp transfer method. The constituent two-dimensional (2D) materials are first exfoliated from bulk crystals onto Si/SiO$_2$ substrates (University Wafer Inc.) at T ~ 60 °C by using silicone-free adhesive film (Ultron Systems Inc.). We separately exfoliate thin layers of graphite (HQ Graphene Inc.), BN, and monolayer WSe$_2$ (HQ Graphene Inc.). Afterward, we apply a polycarbonate-based dry-transfer technique to pick up each layer in series – a top layer of BN, then two thin layers of graphite (as the source and drain contacts), monolayer WSe$_2$, a bottom layer of BN (as the gate dielectric), and another thin layer of graphite (as the back-gate electrode). The entire stack is then deposited on a Si/SiO$_2$ substrate. Throughout the whole transfer process, only the top BN layer contacts the polymer stamp; this reduces the interfacial contaminants and bubbles. Finally, we anneal the devices at T = 300 °C for three hours in a low-pressure argon environment and deposit the electrodes by standard e-beam lithography.

All the experimental results in this paper were obtained in one device. We have measured the heterostructure topography of this device by atomic force microscopy (AFM) to obtain the thickness of the top BN (20 nm), bottom BN (42 nm), and back-gate graphite electrode (9 nm). The SiO$_2$ epilayer is 285 nm in thickness. These thickness parameters, denoted in Fig. S1(c), are used to solve the problem of optical interference to obtain the optical conductivity of monolayer WSe$_2$. The bottom BN thickness (42 nm) is also used to calculate the capacitance of the back gate.

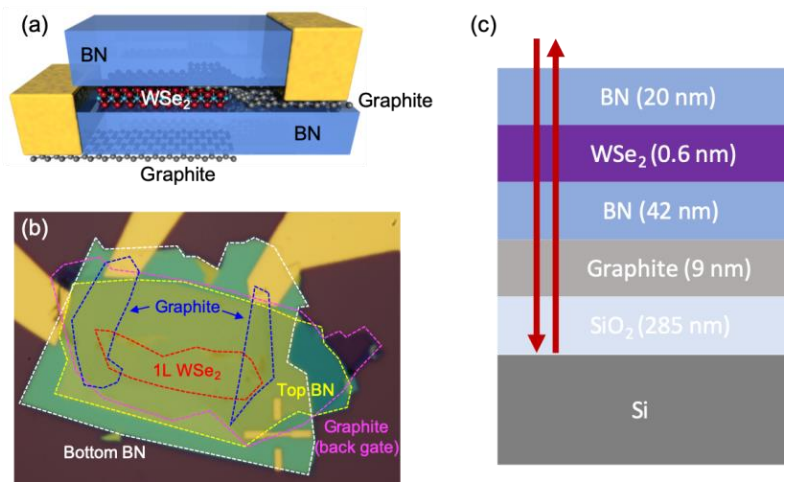

**Figure S1.** (a-b) The schematic and optical image of our BN-encapsulated monolayer WSe$_2$ gating device. (c) The layered structure of our device with the thickness of each material. The arrows denote the propagation direction of light in our experiment.

## 2. Magneto-optical experiments

Our magneto-optical experiments were performed in the SCM3 superconducting DC magneto-optical system in the National High Magnetic Field Laboratory in Tallahassee, Florida, United States. The system has tunable magnetic field from B = –17.5 T to +17.5 T. Fig. S2 displays a schematic experimental setup. The WSe$_2$ device is mounted on an insert inside the cryostat, with the WSe$_2$ plane perpendicular to the magnetic field. The sample temperature is maintained at T ~ 4 K in our experiments. The sample position can be adjusted by a three-axis open-loop piezoelectric positioner. The system uses free-space optics (not fiber optics) to direct the light.

In the reflection experiment, we focus the broadband white light onto the sample with a spot diameter of ~2 μm by an objective lens (NA = 0.6). The reflected light is collected by the same objective and analyzed by a spectrometer (IsoPlane, Princeton Instruments Inc.). Reflected light with right- or left-handed circular polarization is selected by a quarter-wave plate and a linear polarizer. We measure the reflection spectrum on the sample position ($R_s$) and the reference reflection spectrum on a nearby area without the WSe$_2$ monolayer ($R_r$). The fractional change of reflectance is $\Delta R/R = (R_s - R_r)/R_r$. By measuring $\Delta R/R$ at the right- and left-handed circular polarization, we can probe the optical transitions in the K valley and K' valley, respectively.

In the photoluminescence (PL) experiment, we use a 532-nm continuous laser as the excitation source. The laser beam passes through a quarter-wave plate to obtain circular polarization. The laser beam is then focused by an objective lens (NA = 0.6) onto the sample with a spot diameter of ~2 μm. The incident laser power on the sample is estimated to below 10 μW. The PL is collected by the same objective. The PL helicity is selected by a quarter-wave plate and a linear polarizer. Finally, the PL is filtered by a long-pass filter (to remove the excitation light) and analyzed by a spectrometer. In our experiment, we select the same helicity (either right-handed or left-handed) for both the excitation laser and the detected PL.

We measure the gate-dependent reflectance contrast and PL maps of monolayer WSe$_2$ under different magentic fields from B = 0 to 17.5 T. The change of magnetic field, as we note, can cause a slight drift of the sample position. Therefore, every time after we change the magnetic field, we need to adjust the sample back to the original position by tracing the optimized optical signals. As such adjustment is not perfect, the optical spectra can exhibit small irregular changes in energy and spectral profile at different magnetic fields. In our analysis, we focus on the gate-dependent change of optical spectra under the same magnetic field, as well as the gate-voltage positions of half-filled Landau levels (LLs) at increasing magnetic field. These results are not sensitive to the small irregular change of optical spectra caused by the change of magnetic field.

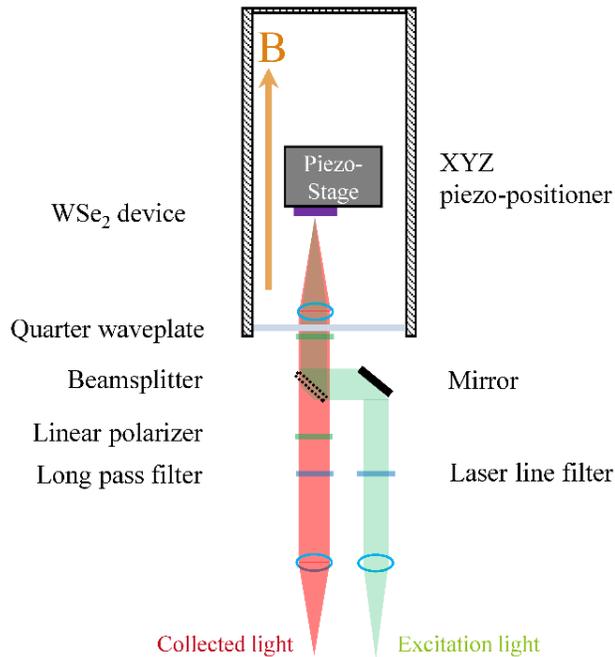

**Figure S2.** Schematic of our magneto-optical setup

## 3. Extraction of optical conductivity from the reflectance contrast

Our measured $\Delta R/R$ reflectance contrast spectra have contribution from both the real and imaginary parts of the conductivity of monolayer $WSe_2$ through the optical interference in the $BN/WSe_2/BN$/graphite/$SiO_2$/Si heterostructure (Fig. S1). We have solved the optical problem in our stacked material system with the transfer matrix method in linear optics [1]. We assume that light propagates as plane waves through the layered structure along the normal direction. We have measured the thickness of the top BN, bottom BN, and graphite electrodes in our device by atomic force microscopy [see the thickness values in Fig. S1(c)]. We adopt the wavelength-dependent complex refractive indices of these materials from the literature [2]. The refractive index of BN is taken to be $n_{BN} \sim 2.13$ for the frequency range of our experiment; it is consistent with the range $2.0 < n_{BN} < 2.3$ generally reported in the literature [3, 4].

Our method of extracting the conductivity is similar to the Kramers–Kronig constrained variational analysis developed by A. B. Kuzmenko [5]. We approximate the spectra of the real and imaginary conductivity by the linear combination of a series of complex Lorentizian functions, each of which follow the Kramers–Kronig relationship. By varying the central energy, width and magnitude of these Lorentizian functions, we optimize the conductivity spectra to match the experimental reflectance contrast spectra. Fig. S3 illustrates a result of our analysis. By using 20 complex Lorentzian functions to approxiate the conductivity spectra, we can closely fit the experimental reflectance contrast spectrum.

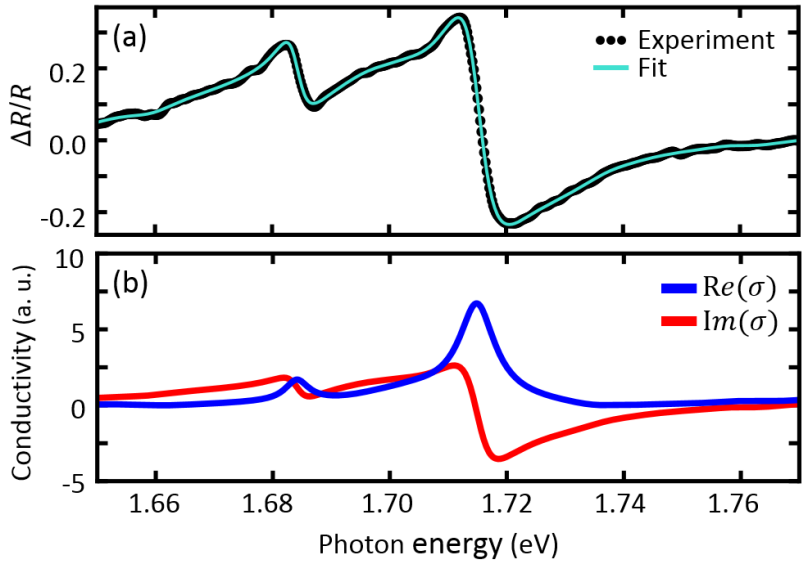

**Figure S3.** (a) An experimental $\Delta R/R$ reflectance contrast spectrum of monolayer $WSe_2$ in the K valley at $V_g = 0.7$ V and B = 17.5 T. The line is a fit by the Kramers–Kronig constrained variational method. (b) The extracted real and imaginary parts of the sheet conductivity spectra for monolayer $WSe_2$.

## 4. Supplemental magneto-reflection results

In Fig. 2-4 of the main paper, we show several representative gate-dependent reflectance contrast and conductivity maps for the K-valley excitonic states. Here we present a more complete set of results in Fig. S4 – S7 for the K-valley optical response under different magnetic fields. We also show the K'-valley reflectance contrast and conductivity maps in Fig. S8.

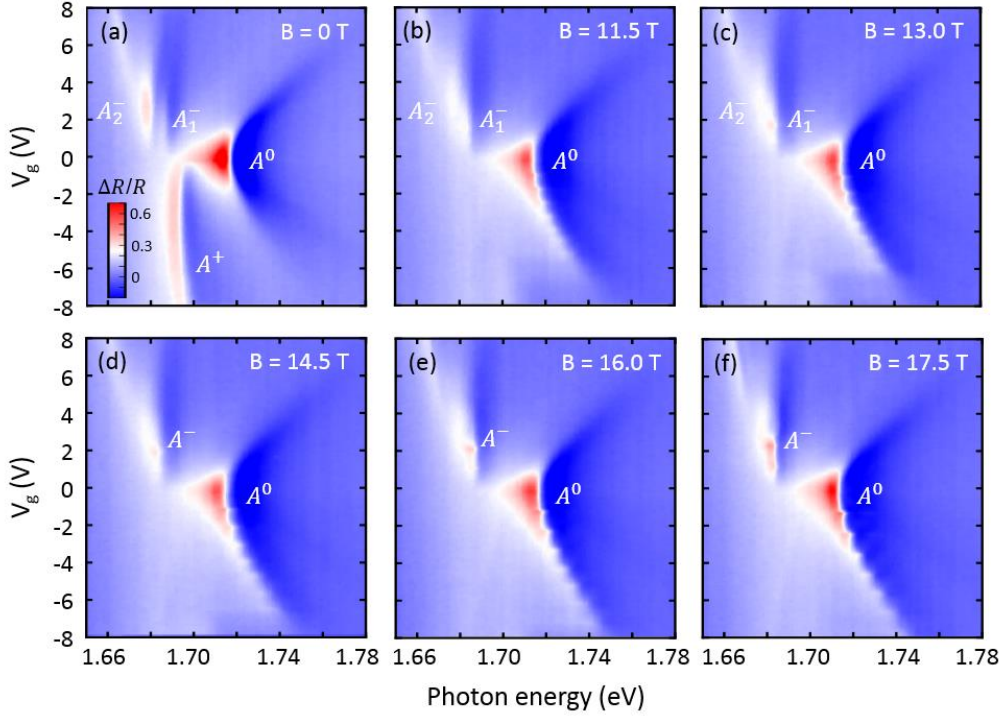

**Figure S4.** (a-f) Gate-dependent reflectance contrast maps for the K-valley excitonic states in monolayer $WSe_2$ under magnetic fields B = 0, 11.5, 13.0, 14.5, 16.0, 17.5 T. We have measured the reflection at the right-handed circular polarization. The experiments were conducted at temperature T ~ 4 K.

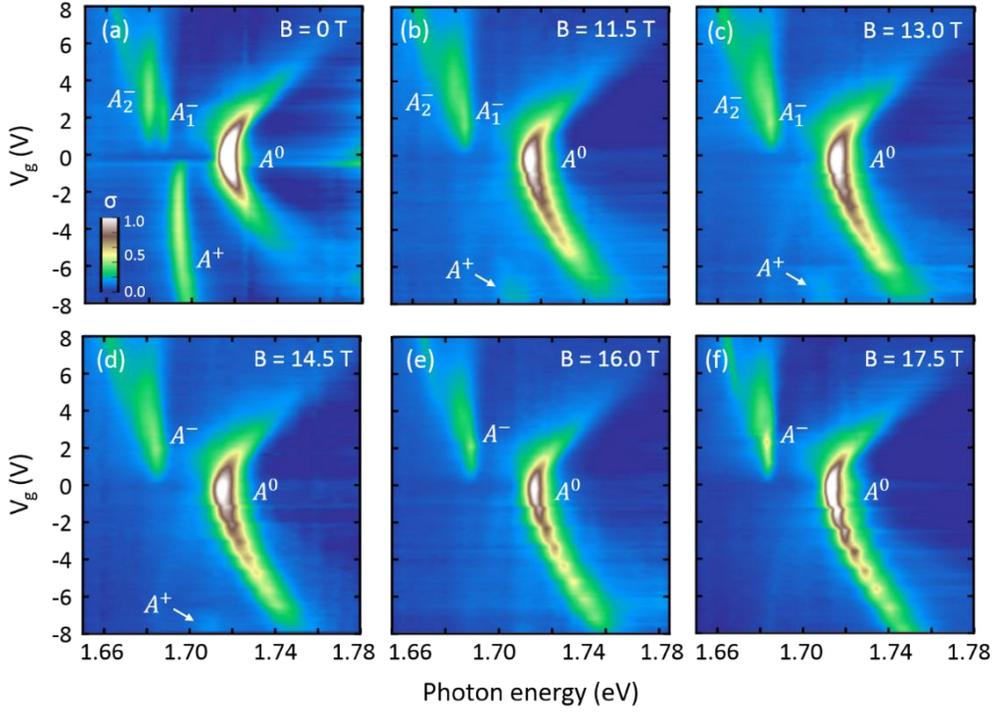

**Figure S5.** (a-f) The real part of the optical sheet conductivity ($\sigma$) for the K-valley excitonic states in monolayer $WSe_2$ as extracted from the reflectance contrast maps in Fig. S4 by using the variational analysis described in the text. The $A^+$ trion is weakly visible at B < 15 T, but totally suppressed at B > 15 T.

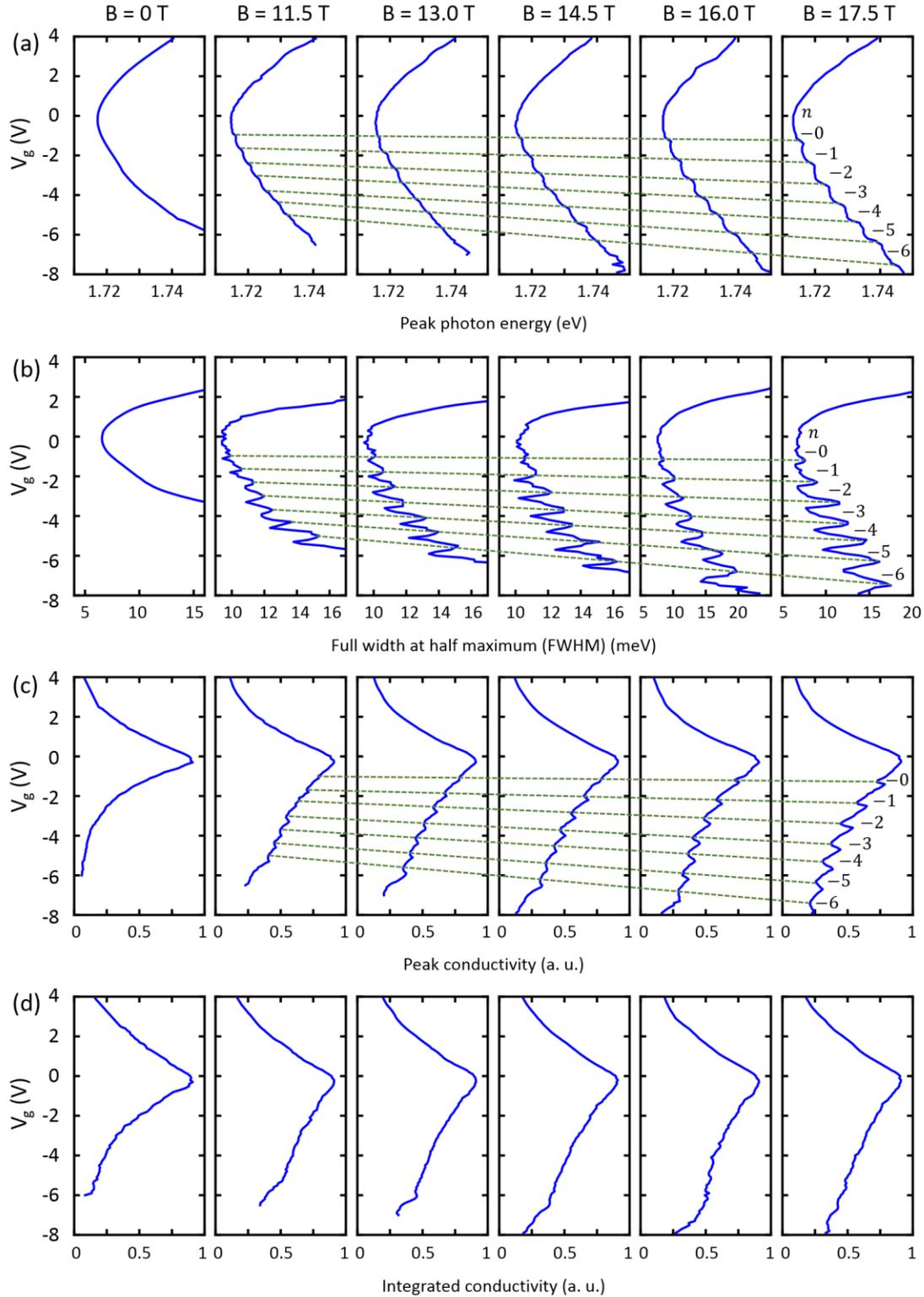

**Figure S6.** (a) The peak energy, (b) full width at half maxium (FWHM), (c) peak conductivity, and (d) integrated conductivity of the K-valley $A^0$ exciton in monolayer WSe$_2$ as a function of gate voltage (y-axis) and magnetic field (columns). The dashed lines denote the half-filled positions of the Landau levels ($n = 0$ to $-6$). The results are extracted from the conductivity maps in Fig. S5.

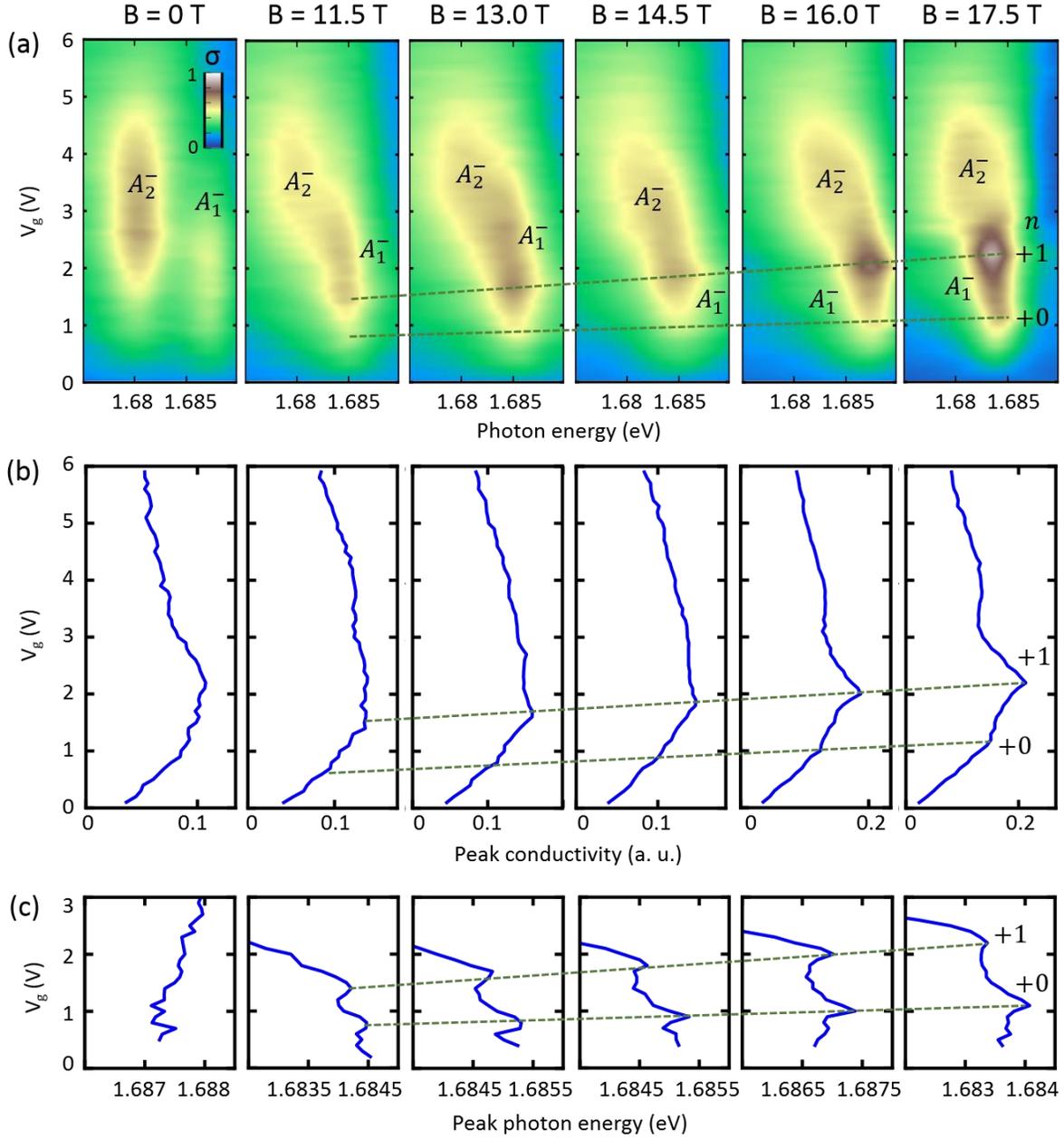

**Figure S7.** (a) The gate-dependent maps of the real part of the optical sheet conductivity ($\sigma$) for the K-valley $A^-$ trions in monolayer WSe$_2$. (b) The peak conductivity magnitude as a function of gate voltage. (c) The peak photon energy, obtained from the Gaussian fitting, as a function of gate voltage. The columns correspond to different magnetic fields. At B = 0 T, the $A_1^-$ intervalley trion is separated from the $A_2^-$ intravalley trion; we analyze only the $A_1^-$ feature at B = 0 T. As the B field increases, the $A_2^-$ feature upshifts in both the gate voltage and energy and merge with the $A_1^-$ feature. We analyze the merged feature (denoted as $A^-$) at B = 11.5 to 17.5 T. The dashed lines denote the half-filled positions of the Landau levels ($n = +0, +1$).

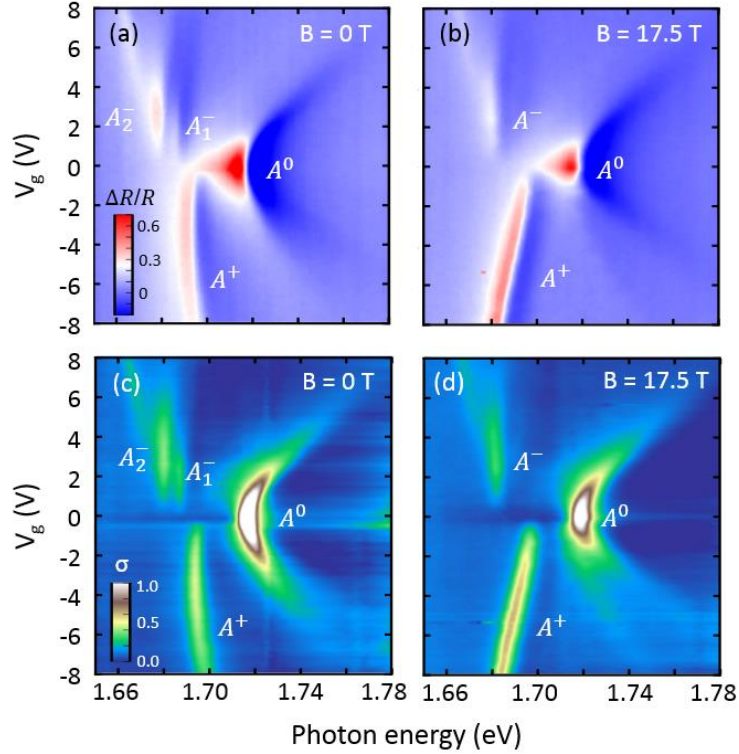

**Figure S8.** (a-b) Gate-dependent reflectance contrast maps for the K'-valley excitonic states in monolayer WSe$_2$ at B = 0 T and 17.5 T. We measure the reflection at left-handed circular polarization to probe the optical transitions in the K' valley. The sample temperature is T ~ 4 K. (c-d) Maps of the real part of the optical sheet conductivity for the K'-valley states, extracted from (a-b).

In Fig. S5, the $A^+$ trion in the K valley is weakly visible at B < 15 T and V$_g$ < –6 V on the hole side, but is completely suppressed at B > 15 T in our measured gate range. This indicates that the Fermi level has not yet reached the valence band of the opposite (K') valley in our maximum gate range at B = 17.5 T. The observed Landau levels (LLs) in the $A^0$ exciton ($n$ = –0 to –6) are all polarized in the K valley (see the schematic band configuration in Fig. 1(b) of the main paper).

In Fig. S7, we observe two trions ($A_1^-$, $A_2^-$) on the electron side. $A_1^-$ is the intervalley trion, formed by coupling an exciton in the K valley and the electron Fermi sea in the K' valley [Fig. 4(d) in the main paper]. $A_2^-$ is the intravalley trion, formed by coupling an exciton in the K valley and the electron Fermi sea in the K valley of the lower conduction band. They are well separated at B = 0 T. As the magnetic field increases, the $A_2^-$ feature upshifts in the gate voltage due to the shift of the valleys, and merges with the $A_1^-$ feature at high magnetic field. At B = 17.5 T, the $A_2^-$ feature appears at higher gate voltges than the two oscillating features of $A_1^-$. This indicates that the two LLs ($n$ = +0, +1) are polarized in the K' valley (see the schematic band configuration in Fig. 1(b) of the main paper)

## 5. Supplemental magneto-photoluminescence results

In Fig. 5 of the main paper, we only show the PL maps of monolayer WSe$_2$ at B = 17.5 T. Here we present a more complete set of PL intensity maps and the associated analysis in Fig. S9 – S12 for both the K-valley and K'-valley emission under different magnetic field strengths.

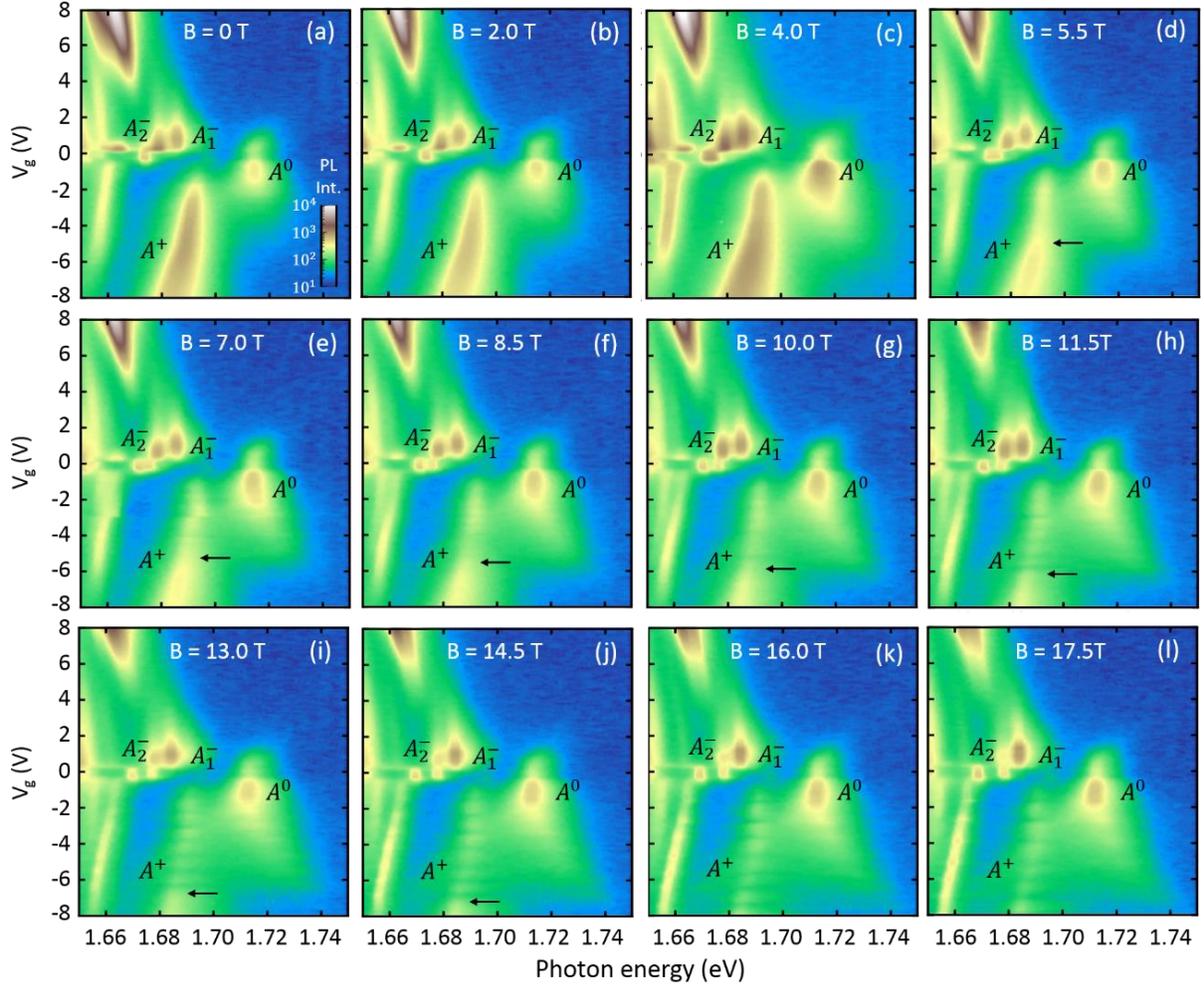

**Figure S9.** (a-l) Gate-dependent right-handed PL maps for the K-valley emission of bright excitonic states in monolayer WSe$_2$ from B = 0 to 17.5 T. The sample temperature is T ~ 4 K. All of the maps share the same logarithmic color scale bar in panel (a). The black arrows denote an abrupt increase of the $A^+$ PL intensity.

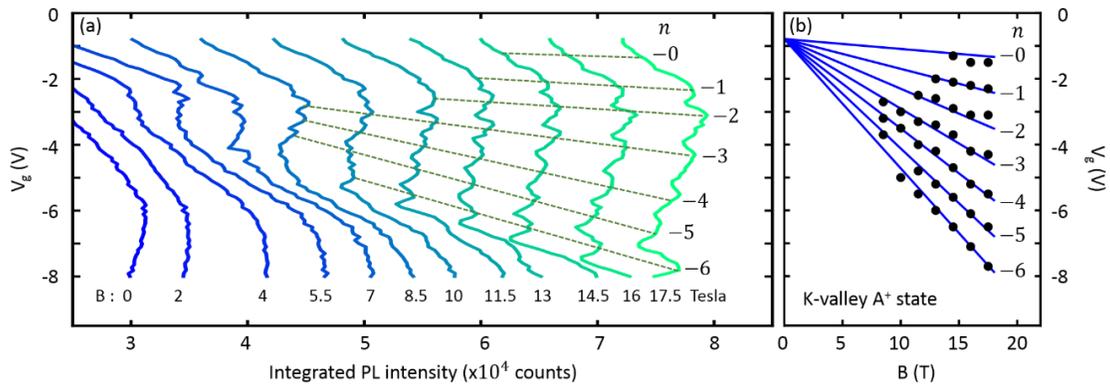

**Figure S10.** (a) The integrated PL intensity of the K-valley $A^+$ state, as extracted from Fig. S9, from B = 0 to 17.5 T. The dashed lines denote the half-filled LL positions ($n$ = –0 to –6) in the K valence valley (see the band configuration in Fig. 5(c) of the main paper). (b) The half-filled LL positions in the K valley extracted from panel (a). The lines are the calculated Landau fan diagram to fit the data.

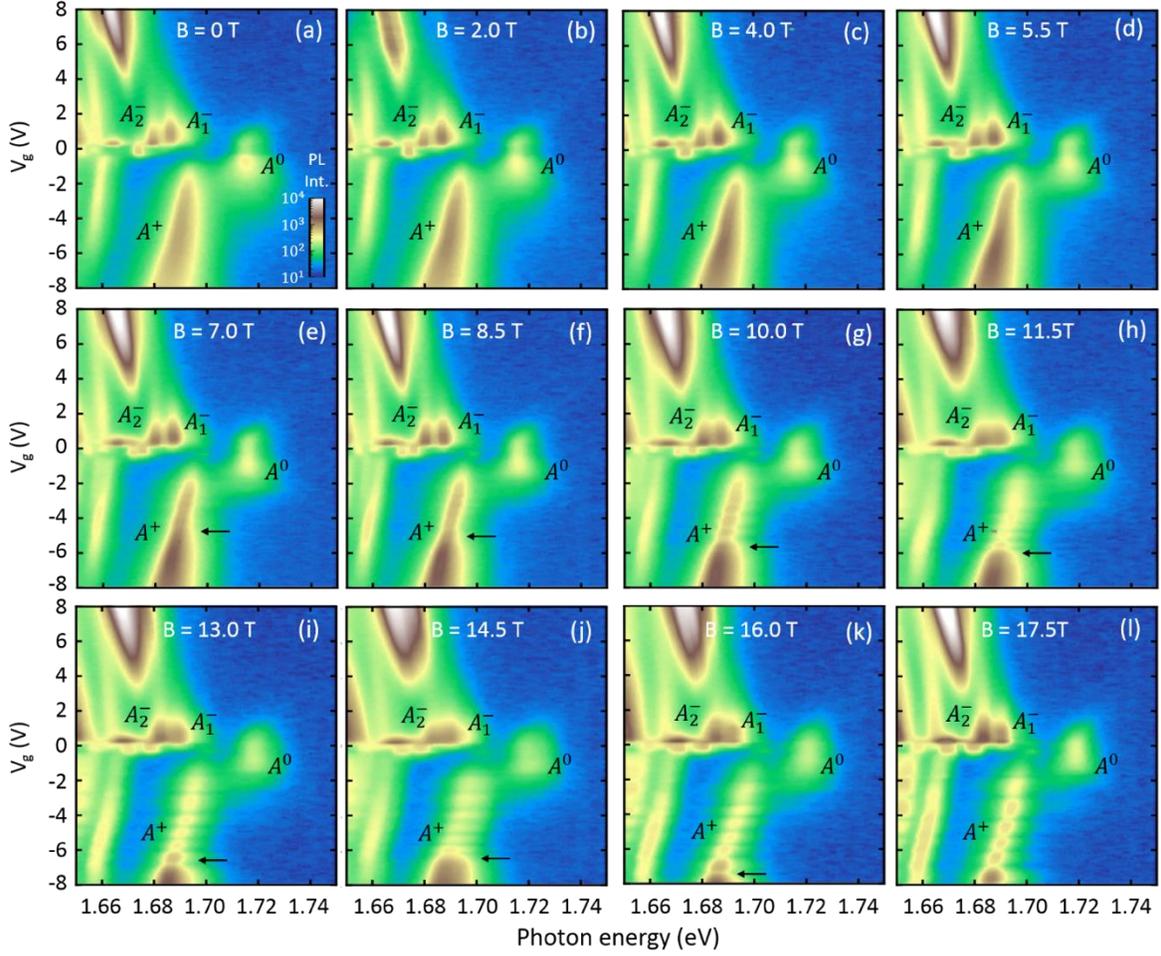

**Figure S11.** (a-l) Gate-dependent left-handed PL maps for the K'-valley emission of bright excitonic states in monolayer WSe$_2$ at B = 0 to 17.5 T. The sample temperature is T ~ 4 K. All of the maps share the same logarithmic color scale bar in panel (a). The black arrows denote an increase of the $A^+$ PL intensity.

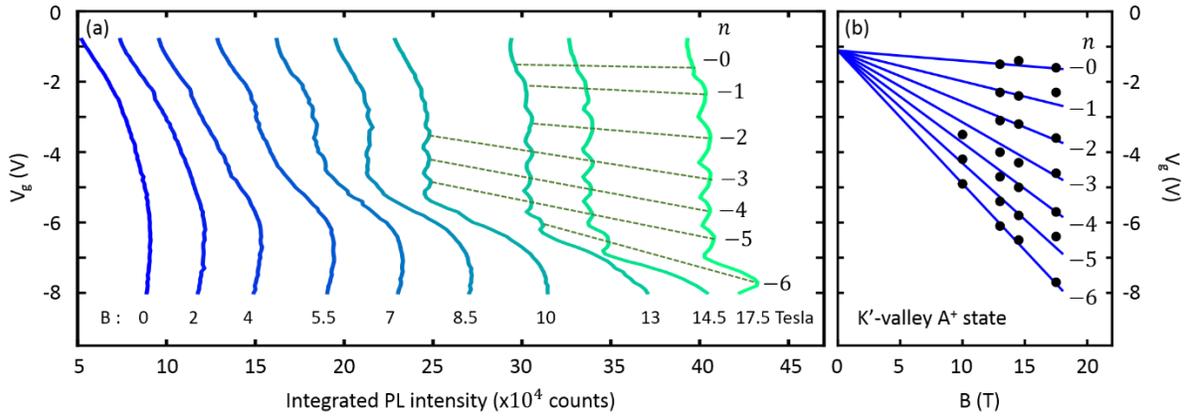

**Figure S12.** (a) The integrated PL intensity of the K'-valley $A^+$ state, as extracted from Fig. S11, from B = 0 to 17.5 T. The dashed lines denote the half-filled LL positions ($n = -0$ to $-6$) in the K-valley valence band (see the band configuration in Fig. 5(f) of the main paper). (b) The half-filled LL positions in the K valley extracted from panel (a). The lines are the calculated Landau fan diagram to fit the data.

We have observed an abrupt increase of the $A^+$ PL intensity at a certain negative gate voltage for both the K-valley and K'-valley emission (marked by the arrows in the Fig. S9 and S11). The PL oscillation is smoothened and obscured after such an increase of PL intensity. We interpret this feature as a signature that the Fermi energy reaches the K' valley after the K valley (see the band configurations in Fig. 5(c, f) of the main paper). The overlapping energy of the states in the two valleys increases the phase space for carrier scattering. This should facilliate the trion formation and enhance the trion PL intensity. The valley-mixing threshold $V_g$ increases with the magnetic field because the valley Zeeman shift increases linearly with the magnetic field. At B = 17.5 T, the threshold $V_g$ goes beyond our measured $V_g$ range. We observe seven LLs ($n = -0$ to $-6$) below the threshold gate voltage; these LLs are all polarized in the K valley.

## 6. Quantum oscillations of dark trions and phonon replicas

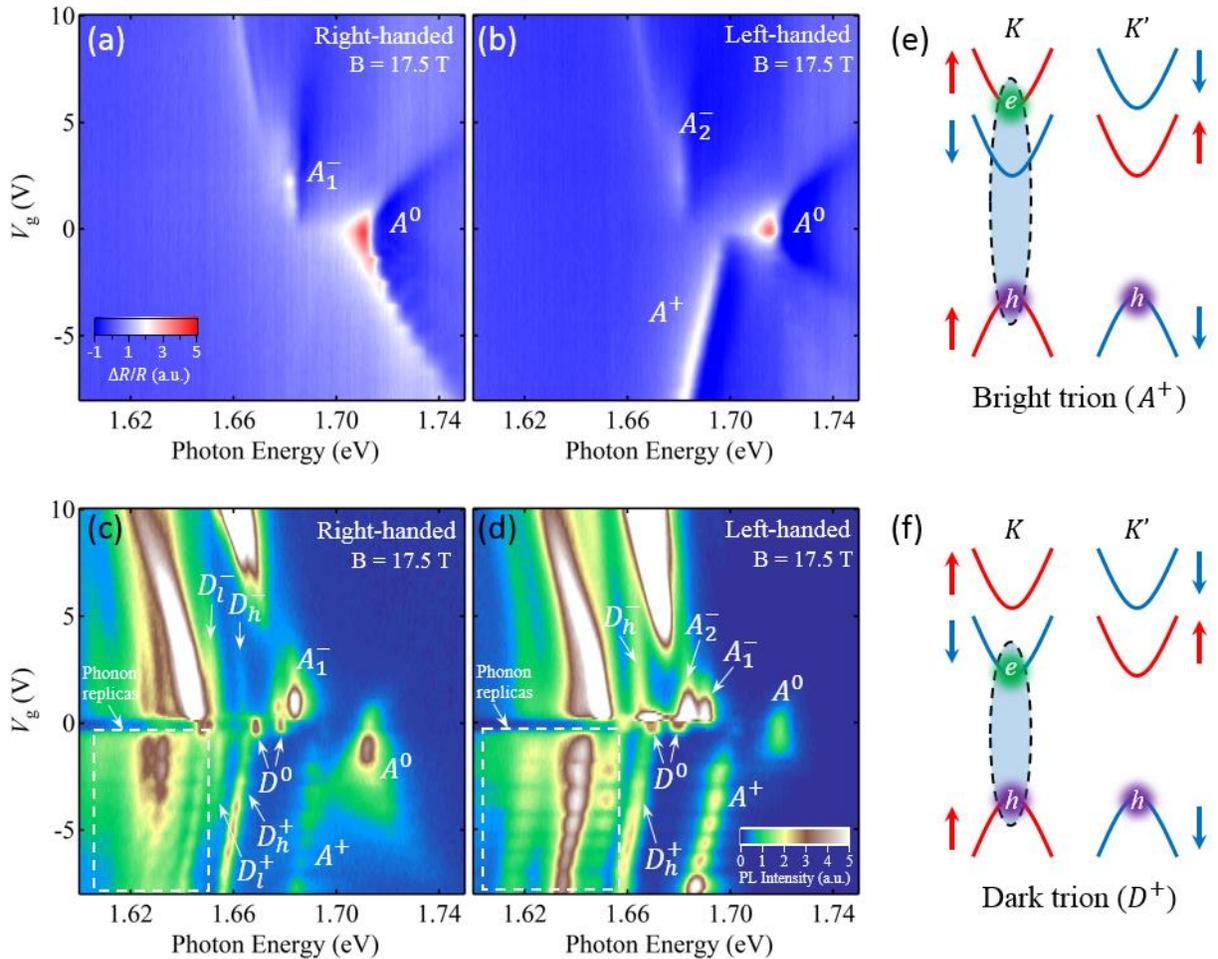

**Figure S13.** (a-b) The gate-dependent reflectance contrast maps at right-handed and left-handed circular polarization (corresponding to optical transitions in the K and K' valleys, respectively) in monolayer WSe$_2$ at B = 17.5 T. (c-d) The gate-dependent PL maps at right-handed and left-handed circular polarization in monolayer WSe$_2$ at B = 17.5 T, corresponding to Fig. 5 (a) and (d) in the main paper, respectively. On the hole side, we observe the quantum oscillations of dark trions and their phonon replicas (dashed box). (e-f) Band configurations of the $A^+$ bright trion and $D^+$ dark trion in monolayer WSe$_2$ with no magnetic field.

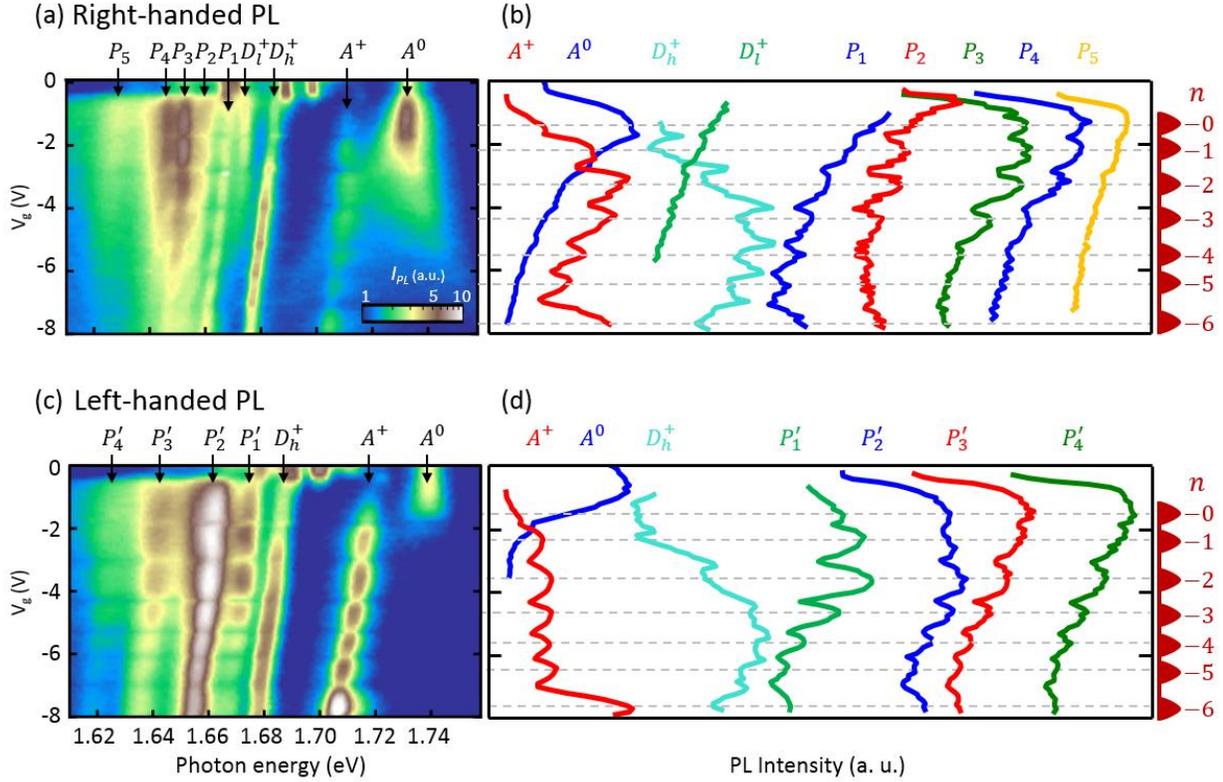

**Figure S14.** (a) The hole-side right-handed PL map from Fig. S13(c) at B = 17.5 T. (b) The integrated PL intensity as a function of gate voltage ($V_g$) for the different peaks in (a). Panels (a) and (b) share the same y-axis of gate voltage. (c-d) Similar as (a-b) for the hole-side left-handed PL map from Fig. S13(d). $D_l^+$ and $D_h^+$ denote the lower-energy and higher-energy PL lines of dark trions, respectively. $P_1 - P_5$ and $P_1' - P_4'$ denote the phonon replicas of dark trions. The dashed lines denote the estimated half-filled LL positions.

We have carried out comparative optical studies on the bright and dark excitonic states under magnetic field [Fig. S13]. In monolayer WSe$_2$, the conduction band is split by spin-orbit coupling into two subbands with opposite spins and small energy separation (~30 meV). They give rise to two kinds of excitons and trions. Those involving the upper conduction band are the bright excitons and trions ($A^0, A^-, A^+$) [Fig. S13(e)]. Those involving the lower conduction band are spin-forbidden dark excitons and trions ($D^0, D^-, D^+$) [Fig. S13(f)]. We do not observe the dark states in the reflectance contrast maps because their oscillator strength is small [Fig. 13(a-b)]. But we can observe the dark states in the PL maps [Fig. 13(c-d)], because the dark states, lying at the lowest energy levels, can accumulate sufficient population to achieve observable luminescence [6].

Unlike the bright excitonic states that emit right-handed (left-handed) light from the K (K') valley, the dark excitons and trions in monolayer WSe$_2$ emit linearly polarized light from either valley. Our circularly-polarized PL maps can therefore show the dark-state emission from both valleys. When an out-of-plane magnetic field is applied to lift the valley degeneracy, the dark-state emission from the K and K' valleys have different energy and hence appear as two distinct lines in our PL maps [Fig. 13(c-d)]. We denote the lower-energy (higher-energy) dark-trion line as $D_l^+$ ($D_h^+$) on the hole side, and $D_l^-$ ($D_h^-$) on the electron side.

The dark-state emission is accompanied by a series of phonon replicas [7, 8] [Fig. S13(c-d)]. Although these phonon replicas originate from the dark states, they acquire oscillator strength from the coupling to the bright states. As a result, they follow the optical selection rules of the bright states and emit circularly polarized light [7, 8]. That is, the phonon replicas exhibit right-handed (left-handed) circular polarization if they arise from coupling to the bright excitonic states in the K (K') valley.

We observe quantum oscillations of dark trions and their phonon replicas on the hole side [Fig. 14]. Fig. 14 (b) and (d) display the integrated PL intensity of the dark trions and their phonon replicas as a function of gate voltage, in comparison with that of the bright exciton and trions. The dark trions and phonon replicas exhibit the same oscillation period as the bright excitonic states, indicating that these oscillations are all induced by the Landau quantization. The lower-energy dark-trion line ($D_l^+$) and the trion replicas ($P_1 - P_5$; $P_1' - P_4'$) oscillate in phase with the $A^+$ bright trion. This is reasonable because partially filled LLs should favor the formation of both bright and dark trions. They should both have maximum PL intensity in half-filled LLs (dashed lines in Fig. 14). Moreover, as the phonon replicas acquire oscillator strength from the coupling to the bright trion state, the replica emission intensity should follow the $A^+$ emission intensity.

The higher-energy dark-trion line ($D_h^+$), however, oscillates inversely to the $A^+$ and $D_l^+$ lines. The $D_h^+$ ($D_l^+$) PL intensity reaches local minima (maxima) at partially filled LLs. We are uncertain about the exact origin for such opposite oscillations. One possible explanation is that $D_l^+$ and $D_h^+$ are competing for the holes in the K' valence valley [Fig. 5(c) and (f) of the main paper]. The formation of an intervalley $D_h^+$ or $D_l^+$ trion requires one hole in the K valley and one hole in the K' valley. While the holes in the K valley are supplied bountifully by gating, the holes in the K' valley are only supplied by the laser excitation [Fig. 5(c) and (f)]. Considering the relaxation of excited holes toward the K valley, the steady-state hole population in the K' valence valley is expected to be small and insufficient to form both the $D_l^+$ and $D_h^+$ trions with all the electrons in the K and K' conduction valley, respecitvely. When the $D_l^+$ trions are efficient to form at partially filled LLs, they will capture many holes in the K' valley. The shortage of holes in the K' valley will suppress the formation of $D_h^+$ trions. Conversely, when the $D_l^+$ trion formation is less efficient at fully filled LLs, more holes in the K' valley will be available to form $D_h^+$ trions. As a result, the $D_h^+$ and $D_l^+$ PL intensity will oscillate inversely with LL filling. Further research is merited to verify this qualitative picture and give quantitiative explanation to the opposite oscillations of dark trions.

## 7. Calculation of the gate-dependent Landau fan diagrams

In Fig. 2(e), S10(b) and S12(b), we use the Landau fan diagrams to match the quantum oscillation data. To construct the diagrams, we assume a constant capacitance ($C = \epsilon_0 \epsilon_{BN}/d$) in our gating device, where $\epsilon_0$ is the permittivity of free space, $\epsilon_{BN}$ is the BN dielectric constant, and $d$ = 42 nm is the measured bottom BN thickness for the back gate. The injected charge carrier density ($N$) is related to the gate voltage as $N = CV_g/e$, where $e$ is the electron charge. We consider the Landau level (LL) degeneracy $N_{LL} = \frac{eB}{2\pi\hbar}$, which excludes the spin and valley degeneracy. The gate voltage needed to half-fill the first LL ($n$ = 0) is:

$$V_{g,n=0} = \frac{1}{2}\frac{eN_{LL}}{C} + V_0. \tag{1}$$

Here $V_0$ is the gate voltage needed to tune the Fermi level to the edge of the conduction band (or valence band) from the charge neutrality point. The $V_0$ value depends on the defect density and Schottky barrier of the device. Since $N_{LL}$ increases linearly with B, we expect the $V_g$ position of the $n = 0$ LL to increase linearly with B as well.

Generally, the gate voltage to half-fill a LL with filling factor $n$ is:

$$V_{g,n} = \left(n + \frac{1}{2}\right)\frac{eN_{LL}}{C} + V_0. \tag{2}$$

This relationship does not depend on the effective carrier mass. We use $\epsilon_{BN}$ and $V_0$ as the fitting parameters to match the quantum oscillation data. For the absorption data in Fig. 2(e), the best fit values are $\epsilon_{BN} = 3.07$ and $V_0 = +0.5\ V$ on the electron side and $V_0 = -0.67\ V$ on the hole side. For the K-valley PL data in Fig. S9(b), the best fit values are $\epsilon_{BN} = 3.03$ and $V_0 = -0.79\ V$. For the K'-valley PL data in Fig. S11(b), the best fit values are $\epsilon_{BN} = 3.14$ and $V_0 = -1.1\ V$. The obtained $\epsilon_{BN} = 3.03 - 3.14$ values are consistent with the BN dielectric constant reported in the literature for relatively thick BN ($d > 30$ nm) [9, 10]. The small variations of $\epsilon_{BN}$ and $V_0$ are presumably due to the different experimental conditions in the measurements, such as the drift of sample position and the different optical excitation conditions for the reflection and PL experiments.

## 8. Estimation of g-factors from the valley-separated Landau levels

Researchers have extensively used the single-particle model to describe the valley Zeeman shift in monolayer transition metal dichalcogenides (TMDs). Despite its simplicity, the model describes well the measured Zeeman shift in low charge carrier density [11-14]. In the single-particle model, the valley Zeeman shift of monolayer WSe$_2$ has three contributions [11, 15]. The first contribution comes from the spin, with energy shift $E_s = 2s\mu_B B$; the second contribution comes from the orbital angular momentum, with energy shift $E_o = m\tau\mu_B B$; the third contribution comes from the self-rotation of the wave packet due to the Berry curvature, with energy shift $E_B = \frac{m_0}{m^*}\tau\mu_B B$. Here $s = \pm\frac{1}{2}$ is the spin quantum number of the band; $\tau$ is the valley index for the K valley ($\tau = +1$) and K' valley ($\tau = -1$); $m$ is the azimuthal quantum number for the atomic orbits in the conduction band ($m = 0$) and valence band ($m = 2$); $\mu_B = 5.8 \times 10^{-5}$ eV/T is the Bohr magneton; $m_0$ is the free electron mass; $m^*$ is the effective carrier mass. The total Zeeman energy shift is the sum of these three contributions:

$$E_Z = \left(2s + m\tau + \frac{m_0}{m^*}\tau\right)\mu_B B = g\mu_B B. \tag{3}$$

Here $g = 2s + m\tau + \frac{m_0}{m^*}\tau$ is the total effective g-factor of the band. We use the effective mass $m^* = 0.4m_0$ for both the electron and hole, according to the values in the literature [16-18]. In a simple estimation for the K valley ($\tau = +1$), $g = 3.5$ for the upper conduction band ($s = +\frac{1}{2}$, $m = 0$), $g = 1.5$ for the lower conduction band ($s = -\frac{1}{2}$, $m = 0$), and $g = 5.5$ for the upper valence band ($s = +\frac{1}{2}$, $m = 2$). The corresponding bands in the K' valley have the same g-factor with negative sign due to the time reversal symmetry.

In our analysis here, we define a LL to be valley-separated when it lies below (above) the lowest (highest) LL in the same conduction (valence) band of the opposite valley (*i.e.* it is out of the energy range of the opposite valley; see Fig. 1 of the main paper). In the single-particle model, we can calculate the number of valley-separated LLs from the predicted g-factors and the LL energy spacing $\Delta E_{LL} = 2\frac{m_0}{m^*}\mu_B B \cong 5\mu_B B$. The energy difference of the upper conduction band ($c_2$; $g = \pm 3.5$) between the K and K' valley is $\Delta E_{c2} = 7\mu_B B > \Delta E_{LL}$; there are two valley-separated LLs ($n = +0, +1$) in the upper conduction band. Similarly, the energy difference of the lower conduction band ($c_1$; $g = \pm 1.5$) between the K and K' valley is $\Delta E_{c1} = 3\mu_B B < \Delta E_{LL}$; there is only one valley-separated LL ($n = +0$) in the lower conduction band. The energy difference of the valence band ($v_1$; $g = \pm 5.5$) between the K and K' valley is $\Delta E_{v1} = 11\mu_B B > 2\Delta E_{LL}$; there are three valley-separated LLs ($n = -0, -1, -2$) in the valence band. We summarize these results in Table S1 and also in Fig. 1(a) of the main paper. We note that the number of valley-separated LLs remain the same even if we use a different electron effective mass ($m^* = 0.1m_0 - 0.9m_0$) or a different hole effective mass ($m^* = 0.35m_0 - 0.9m_0$).

Our experimental results, obtained in relatively high charge density, are very different from the prediction of the single-particle model, which is known to be good in low charge density. In our experiment, we observed two valley-separated LLs ($n = +0, +1$) in the lower conduction band ($c_1$) and seven valley-separated LLs ($n = -0$ to $-6$) in the upper valence band ($v_1$). These numbers are considerably larger than the prediction of the single-particle model. By using the effective mass $m^* = 0.4m_0$ for both the electron and hole, our experimental results give the g-factors $g \approx 2.5$ for the lower conduction band and $g \approx 15$ for the upper valence band. These g-factors are much larger than the prediction of the single-particle model for the lower conduction band ($g = 1.5$) and upper valence band ($g = 5.5$). The enhancement of g-factors are presumably induced by the strong many-body interactions in monolayer WSe$_2$. Recent experiments also report interaction-driven g-factor enhancement in the conduction band of bilayer MoS$_2$ [19], monolayer and bilayer MoSe$_2$ [20] and the valence band of monolayer WSe$_2$ [21]. Here we further demonstrate that the interaction-driven g-factor enhancement exists on both the electron and hole sides of monolayer WSe$_2$.

|  | Single-particle model | | Experimental results | |
| --- | --- | --- | --- | --- |
|  | g-factor | Valley-separated LLs | g-factor | Valley-separated LLs |
| Upper conduction band ($c_2$) | ±3.5 | $n = +0, +1$ | — | — |
| Lower conduction band ($c_1$) | ±1.5 | $n = +0$ | ±2.5 | $n = +0, +1$ |
| Upper valence band ($v_1$) | ±5.5 | $n = -0, -1, -2$ | ±17.5 | $n = -0$ to $-6$ |

**Table S1.** The g-factors and valley-separated Landau levels in the different bands of monolayer WSe$_2$ according to the single-particle model and our experimental results. The associated band configurations are shown in Fig. 1 of the main paper.

## 9. Estimation of the Wigner-Seitz radius at the lowest Landau levels

Our observation of the $n = \pm 0$ LLs allow us to characterize a very low free charge density in both the electron and hole sides of monolayer WSe$_2$. This may enable the realization of Wigner crystallization. We may estimate the Wigner-Seitz radius $r_s$, which measures the average inter-particle separation in units of the effective Bohr radius. $r_s$ characterizes the ratio of electron Coulomb interaction strength to its kinetic energy. It has the form: [22]

$$r_s = \frac{1}{\sqrt{\pi N}} \frac{m^* e^2}{4\pi\varepsilon\varepsilon_0 \hbar^2}. \tag{4}$$

The formula includes the carrier density ($N$) and effective mass ($m^*$), the background dielectric constant ($\varepsilon$) and vacuum permittivity ($\varepsilon_0$), the elementary charge ($e$) and the Planck constant ($\hbar$). Here we consider no spin or valley degeneracy, because only one valley with a specific spin is filled under magnetic field in our experiment.

For a simple estimation of $r_s$, we use the effective mass $m^* \sim 0.4 m_0$ ($m_0$ is the free electron mass) for both the electrons and holes in monolayer WSe$_2$. In our device geometry, the surrounding dielectric medium BN has an effective dielectric constant of $\varepsilon = \sqrt{\varepsilon_\| \varepsilon_\perp}$ with the in-plane and out-of-plane dielectric constants $\varepsilon_\perp = 3.04$ and $\varepsilon_\| = 6.93$ [10]. In our experiment, we observe the half-filled $n = 0$ LLs at B = 11.5 T, which allow us to characterize a very low density $N \approx 1.4 \times 10^{11}$ $cm^{-2}$ for the free carriers (not trapped by the defects).

From the above parameters, we obtain a Wigner-Seitz radius $r_s \approx 25$. This value is close to the predicted Wigner crystallization condition $r_s \gtrsim 31$ in two dimensions [23-26]. Therefore, our experimental results indicate strong electron-electron interactions, which may lead to Wigner crystallization of electrons, in monolayer WSe$_2$. Such interaction conditions cannot be found in graphene, which has small $r_s$ due to the small carrier effective mass.

## 10. Theoretical calculation of exciton quantum oscillation in monolayer WSe$_2$

We have caculated the $A^0$ exciton conductivity spectra of monolayer WSe$_2$ at different charge density under magnetic field by using semi-emiprical calculations based on the effective-mass approximation. The results are shown in Fig. 3(e-h) of the main paper. Here we will provide the details of our calculations.

### 10.1. Free carriers under magnetic field

The band structure of monolayer WSe$_2$ near the K (or K$'$) point is nearly parabolic for both the conduction and valence bands [18, 27, 28]. The electron and hole effective masses are estimated to be around $0.38 m_0$ and $0.44 m_0$ ($m_0$ is the free electron mass), respectively, according to the average values of various calculations based on the density-functional theory (DFT) [18, 27-29].

Under a magnetic field B along the $z$ axis (normal to the plane of monolayer WSe$_2$), a free carrier in monolayer WSe$_2$ can be described by a 2D massive Dirac Hamiltonian [25]

$$H_1 = v_F[\tau(p_x - qA_x)\sigma_x + (p_y - qA_y)\sigma_y] + \Delta\sigma_z - \mu_B B(m\tau + g_s s). \tag{5}$$

Here $v_F$ is the Fermi velocity; $p_x$ and $p_y$ are the momentum operators; $\sigma$'s are the pseudospin Pauli matrices; $q = \pm e$ is the carrier charge; $\Delta$ is a parameter that produces a $2\Delta$ band gap at zero magnetic field; $\tau$ is the valley index for the K valley ($\tau = +1$) and K$'$ valley ($\tau = -1$); $\boldsymbol{A}$ is the vector potential; $m\tau$ and $s = \pm\frac{1}{2}$ are the spin and orbital azimuthal angular momentum quantum number of the band, respectively; and $g_s = 2$ is the spin g-factor. After digonalizing the $2 \times 2$ Hamiltonian and keeping the low-order terms, we get an effective-mass Hamiltonian:

$$H_1 = \frac{1}{2m^*}(\boldsymbol{p} - q\boldsymbol{A})^2 - \mu_B B(m\tau + g_s s) - \frac{1}{2}\tau\hbar\omega_c$$
$$= \frac{p_x^2}{2m^*} + \frac{1}{2m^*}(p_y - qBx)^2 - E_Z. \tag{6}$$

Here $m^* = \tau\Delta/v^2$ is the effective carrier mass; $\omega_c = \frac{eB}{m^*}$ is the cyclotron frequency; and $E_Z$ is the net Zeeman shift. Here we have used the Landau gauge with $\mathbf{A} = Bx\hat{\mathbf{y}}$ [26]. The last term in Eq. (6) comes from the intrinsic orbital magnetic moment induced by the pseudospin-orbit coupling [30]. Since the system is translationally invariant along $y$, we can replace $p_y$ by $\hbar k_y$ with $k_y$ being a good quantum number. Eq.(6) can be rewritten as:

$$H_1 = \frac{p_x^2}{2m^*} + \frac{1}{2}m^*\omega_c^2(x-x_0)^2 - E_Z, \tag{7}$$

where $x_0 = \frac{\hbar k_y}{m^*\omega_c}$. The eigenvalues of $H_1$ are simply those for a harmonic oscillator, $E_n = \left(n+\frac{1}{2}\right)\hbar\omega_c - E_Z$. The degeneracy of each energy level is $\frac{L^2}{2\pi\alpha^2}$ [26]. Here $n = 0, 1, 2 ...$ is the quantum number; $L^2$ is the sample area; and $\alpha = \sqrt{\frac{\hbar}{eB}}$ is the radius of cyclotron orbit. The energy levels ($E_n$) are called the Landau levels (LLs). Due to the presence of defects in the sample, these LLs will be broadened by the scattering between carriers and defects. To describe such broadening, we adopt the model of Ref. [31], in which the self-energy for the $n$-th LL is given by:

$$\Sigma_n(E) = \frac{1}{4}\sum_{n'} \Gamma_{nn'}^2 G_{n'}(E). \tag{8}$$

Here $G_n(E) = 1/[E - E_n - \Sigma_n(E)]$ denotes the Green's function and

$$\frac{1}{4}\Gamma_{nn'}^2 = \frac{1}{(2\pi)^2}\int |u(\mathbf{q})|^2 J_{nn'}(\alpha q)d\mathbf{q}. \tag{9}$$

$u(\mathbf{q})$ is a short-range potential describing the defect scattering. $J_{nn'}$ has the expression:

$$J_{nn'}(x) = \sqrt{\frac{n!}{n'!}}\left(\frac{x}{\sqrt{2}}\right)^{n'-n} L_n^{n'-n}\left(\frac{x^2}{2}\right)\exp\left(-\frac{x^2}{4}\right), \tag{10}$$

where $L_n^{n'}(x)$ is the associated Laguerre polynomial. When the $n \neq n'$ terms are neglected in Eq. (8), we have:

$$\Sigma_n(E) \to \Sigma_n^0(E) = [E - E_n - \sqrt{(E-E_n)^2 - \Gamma_{nn}^2}]/2, \tag{11}$$

where $\sqrt{x} = i\sqrt{|x|}$ when $x < 0$. Here we simply take the constant $\Gamma_{nn}$ as an empirical parameter. The density of states of the $n$-th LL (excluding the spin and valley degeneracy) is:

$$D_n(E) = \frac{1}{(\pi\alpha\Gamma_{nn})^2}\sqrt{\Gamma_{nn}^2 - (E-E_n)^2} \tag{12}$$

if $|E - E_n| < \Gamma_{nn}$. The density of states is zero otherwise. Here $\Gamma_{nn}$ is the half-width of the broadening for the $n$-th LL. The carrier concentration is given by:

$$N = \sum_n \int_0^{E_F} dE\, D_n(E) = n'N_0 + N_{n'}(E_F). \tag{13}$$

Here $N_0 = \frac{1}{2\pi\alpha^2}$ is the carrier density in each filled LL. The carrier density in the partially filled LL $n'$ is:

$$\begin{aligned}
N_{n'}(E_F) &= \frac{1}{(\pi\alpha\Gamma_{nn})^2}\int_{E_{n'}-\Gamma_{nn}}^{E_F} dE\,\sqrt{\Gamma_{nn}^2 - (E-E_{n'})^2} \\
&= \frac{1}{(\pi\alpha)^2}\int_{-1}^{u_F} du\,\sqrt{1-u^2} \\
&= \frac{1}{(\pi\alpha)^2}\int_{-\pi/2}^{\theta_F} d(\sin\theta)\cos\theta \\
&= \frac{1}{2(\pi\alpha)^2}\left(\theta_F + \frac{\pi}{2} - u_F\sqrt{1-u_F^2}\right),
\end{aligned} \tag{14}$$

where $u_F = (E_F - E_{n'})/\Gamma_{nn}$ and $\theta_F = \sin^{-1}(u_F) < \pi/2$.

Once the carrier density is determined by Eq. (14) for a given Fermi energy $E_F$, we can obtain an effective $k_F$ according to $k_F = \sqrt{4\pi N}$, which is used to determine the basis functions for constructing the exciton wavefunction and estimate the band renormalization energy. We adopt the following expression of band renormalization energy [32, 33]:

$$\Delta E_{BR} = \int_0^{k_F} \frac{qdq}{2\pi} V(q)/\tilde{\varepsilon}(q) + \int_0^{\infty} \frac{kdk}{2\pi} V(k)[1 - 1/\tilde{\varepsilon}(q)] \tag{15}$$

with $\tilde{\varepsilon}(q) = 1 - \eta V(q) \Pi(q,0)$. (16)

Here the intra-LL polarization $\Pi(q,0)$ is given by the equations (22) and (23) in Ref. [31]. $V(q)$ is the effective Coulomb interaction in monolayer WSe$_2$, which is given by: [34-39]

$$V(q) = -\frac{e^2}{2L^2 \kappa_0 \varepsilon_0 q (1+q\rho_0)}. \tag{17}$$

Here $\kappa_0$ is the effective static dielectric constant of the 2D material; $\rho_0$ is an empirical parameter related to the finite thickness of the 2D material and the screening length of the $q$-dependent dielectric screening [40]. The values of these parameters used here are the same as those in Ref. [41]. The $\eta$ factor in Eq. (16) denotes the spin degeneracy multiplied by a reduction factor for a 2D material encapsulated by undoped materials. Since a large fraction of the field lines for the electrostatic interaction between two charges go out of the 2D plane, the free carrier screening is significantly reduced.

The first term in Eq. (15) describes the exchange interaction of a carrier with other carriers in the same band. Since $\Pi(q,0) = 0$ when the LLs are fully occupied or empty, the correction of band renormalization energy becomes artificially large as the inter-LL interactions are neglected in the current model. In addition, the correlation effect can be siginificant in this low-dimensional system. To remedy this, we adopt a semi-empirical approach. Instead of using only the intra-LL screening $\tilde{\varepsilon}(q)$, we replace it with the dielectric screening of a 2D electron gas [42] with the polarization part modified by a scaling parameter ($\eta'$):

$$\varepsilon(q) = 1 + \eta' V(q) \frac{2\varepsilon_0 m_e}{\hbar^2} (1 - \sqrt{1-(2k_F/q)^2}) \Theta(q - 2k_F). \tag{18}$$

The scaling factor $\eta'$ is treated as an empirical parameter to fit the experimental data. It accounts for the effect of correlation and the deviation of the quasi-2D system from the ideal 2D system. The best-fit value for the current monolayer WSe$_2$ sample is $\eta' = 0.27$.

The second term in Eq. (15) describes the shift of the band edges of both the filled and unfilled bands due to the Coulomb hole [32]. This term only contributes when the LL is partially filled (*i.e.* when $E_F$ lies inside the broadened zone of a LL).

## 10.2. Exciton under magnetic field

We next consider the exciton under magnetic field. We use an exciton Hamiltonian:

$$H_X = \frac{1}{2m_e^*}(\mathbf{p} + e\mathbf{A})^2 + \frac{1}{2m_h^*}(\mathbf{p} - e\mathbf{A})^2 - V(\mathbf{r}_e - \mathbf{r}_h) - \Delta E_Z. \tag{19}$$

Here $V$ is the electron-hole Coulomb interaction whose Fourier transform is given in Eq. (17). $\Delta E_Z$ is the difference in Zeeman shift between the conduction and valence bands. The Coulomb interaction is screened by $\tilde{\varepsilon}(q)$ in Eq. (16) when the LLs are partially filled.

In the symmetric gauge, we have $\boldsymbol{A} = \boldsymbol{B} \times \boldsymbol{r}/2 = (-y, x)B/2$ and Eq. (19) becomes [43]:

$$H_X = \frac{1}{2\mu}(\boldsymbol{p} + e\boldsymbol{A})^2 + V - \Delta E_Z$$
$$= \frac{1}{2\mu}\left(p^2 + e\boldsymbol{B} \cdot \boldsymbol{L} + \frac{e^2}{4}r^2 B^2\right) + V - \Delta E_Z. \quad (20)$$

Here $\mu$ is the reduced mass of the exciton in monolayer WSe$_2$, which has been experimentally measured to be $\mu \approx 0.2 m_0$ ($m_0$ is the free electron mass) [44]. For the $s$-like ground state, we have $\boldsymbol{B} \cdot \boldsymbol{L} = 0$. We adopt an effective dielectric constant $\kappa = 3.97$. The corresponding atomic unit for exciton energy is $R_X = (\mu/\kappa^2)13.6$ eV = 0.172 eV. The atomic unit for distance is $a_X = (\kappa/\mu)0.529$ Å = 10.5 Å.

The exciton state in the presence of a Fermi sea is:

$$|\Psi_X\rangle = \sum_k \varphi_X(\boldsymbol{k}) a_k^+ b_{-k}^+ |F\rangle. \quad (21)$$

Here $\varphi_X(\boldsymbol{k})$ is the exciton envelope function in the k-space; $a_k^+$ ($b_{-k}^+$) are operators creating an electron (hole) in the conduction (valence) band; $|F\rangle$ denotes a Fermi sea of gate-injected carriers under the magnetic field. The exciton states under magnetic field are solved in the same way as in Ref. [41], except here we also include the Pauli blocking effect of the Fermi sea. Since the exciton binding is much larger than the cyclotron energy $\hbar\omega_c$, the Landau quantization enters mainly through the diamagnetic term $\frac{e^2}{8\mu}r^2 B^2$ in Eq. (20) and the quantized screening $\tilde{\varepsilon}(q)$ in Eq. (16). The Pauli blocking effect can be included by imposing a constraint of step function $\Theta(k - k_F)$ on the basis functions used in the expansion of $\varphi_X(\boldsymbol{k})$ in the k-space. The k-space representation of the basis states adopted here is the same as that described in Ref. [45]. The effective Fermi wave vector is related to the carrier density ($N$) by $k_F = \sqrt{4\pi N}$ as discussed in the previous section.

The calculated exciton peak position as a function of gate voltage ($V_g$) is shown in Fig. S15(a). We note that, when $V_g$ is applied, the injected carriers will first fill the defect trapping states inside the band gap of the material until it reaches a threshold gate voltage $V_0$ ($V_0 = +0.5$ V and $-0.67$ V in our absorption data). These initial filling charges will cause a blue shift of exciton energy. The gate-induced vertical electric field will also reduce the exciton oscillator strength. The energy shift during this initial charging process is assumed to be linear in $|V_g|$ with a slope of 0.002 to match the experimental data. After $|V_g|$ reaches $V_0$, the injected carriers begin to fill the lowest LL. When a LL is partially filled, the free-carrier screening can significantly reduce the exciton binding energy. However, this reduction is nearly cancelled by the Coulomb-hole contribution of the band renormalization energy as given in the second term of Eq. (15). When we use the same reduction factor $\eta \sim 0.02$ in $\tilde{\varepsilon}(q)$ both for the carrier screening of the exciton and for the Coulomb-hole contribution in the band renormalization energy, we get a small dip at the half-filled LL positions [blue curve in Fig. S15(a)]. But if we assume that the reduction factor $\eta$ for the Coulomb-hole contribution is scaled by 0.8, a small bump appears at half filling; this matches the experiment better [red curve in Fig. S15(a)].

The broadening of exciton peak due to the intra-LL carrier-carrier scattering in a partially filled LL ($n$) can be approximately described by

$$\text{Im}\,\Sigma_{n,X}(E_F) = \text{Im}\,G_n^0(E_F) \sum_{k,k'} |\langle\varphi_X|V|\varphi_X; k, k'\rangle|^2 \,\Theta(E_F - \varepsilon_n(k))\Theta(\varepsilon_n(k') - E_F). \quad (22)$$

Here $k$ and $k'$ denote the quantum numbers of the states in the $n$-th LL with energies $\varepsilon_n(k)$ and $\varepsilon_n(k')$ which are below and above $E_F$, respectively. $\Theta(x)$ is a step function that imposes the constraint on the occupied or unoccupied states. $V$ is the screened Coulomb potential. $\varphi_X$ denotes the exciton state, which is scattered by $V$ into an exciton accompanying an electron-hole pair (labeled by $k$ and $k'$) in the partially filled LL. In the contact-potential approximation similar to that used in Ref. [33], we have:

$$\text{Im}\,\Sigma_{n,X}(E_F) = \text{Im}\,G_n^0(E_F)|U_n a_X^2|^2 \int_{E_F-\Gamma_{nn}}^{E_F} d\varepsilon D_n(\varepsilon) \int_{E_F}^{E_n+\Gamma_{nn}} d\varepsilon' D_n(\varepsilon')$$
$$= \text{Im}\,G_n^0(E_F)|U_n a_X^2|^2 N_n(E_F)[N_0 - N_n(E_F)]. \quad (23)$$

Here $U_n$ is the strength of the contact potential. We use an empirical form:

$$U_n = U_\infty \sqrt{1 - e^{-0.22(n+0.5)}}, \text{ with } U_\infty = 3.22 R_X \quad (24)$$

to match the local maxima of the observed FWHM (excluding the background FWHM). It is found that the intra-LL scattering is weaker for small $n$. This suggests that carriers in the lower LLs (with smaller cyclotron orbits) are less likely to scatter with the exciton.

The background FWHM is attributed to the inter-LL carrier-carrier scattering, which can be approximated by:

$$\text{Im}\,\tilde{\Sigma}_X(E_F) = \text{Im}\,\sum_{k,k'} G^0(E_F; k, k') |\langle \varphi_X | V | \varphi_X; k, k' \rangle|^2 \, \Theta(E_F - \varepsilon(k))\Theta(\varepsilon(k') - E_F). \quad (25)$$

Here $\varepsilon(k)$ and $\varepsilon(k')$ are the free-carrier band energies at $k$ and $k'$, which are below and above $E_F$, respectively. Here the Landau quantization is neglected. Such a contribution can be evaluated in the contact-potential model as described in Ref. [33] when the coupling of an exciton with the Fermi-sea polarization is considered. According to the findings of Ref. [33], Im $\tilde{\Sigma}_X(E_F)$ increases approximately linearly with the Fermi level $E_F$ [See Fig. 6(b) in Ref. [33]]. Thus, we can write approximately:

$$\text{Im}\,\tilde{\Sigma}_X(E_F) = \Gamma_0 + \beta E_F. \quad (26)$$

Here $\Gamma_0$ is the inhomogeneous line width (~6 meV) of the exciton at zero carrier density; $\beta = 0.25$ is the slope for negative $V_g$. Both $\Gamma_0$ and $\beta$ are empirical parameters to fit our data. We note that in Ref. [33], the exciton is coupled to a Fermi sea with opposite spin with no Pauli exclusion, so the matrix element $\langle \varphi_X | V | \varphi_X; k, k' \rangle$ contains only the direct Coulomb interaction (no exchange interaction); this gives a slope $\beta \approx 0.88$. In our case of the $A^0$ exciton, however, the exciton is coupled to the Fermi sea in the same valley with the same spin, so the exchange Coulomb interaction must be included. This will reduce the overall interaction strength and give a smaller slope $\beta$.

The calculated results for the full width half maximum (FWHM) is shown as the red line in Fig. S15(b). We note that the bumps in FWHM obtained theoretically appear too narrow compared to experiment. This is because we used an energy-independent defect scattering width $\Gamma_{nn}$. In reality, $\Gamma_{nn}$ should be a function of energy and the density of states $D_n(\varepsilon)$ should become more Gaussian-like instead of an abrupt bump as described by Eq. (12). If we replace $D_n(\varepsilon)$ by a Gaussian function in Eq. (23), the simulation results could be improved.

The exciton oscillator strength is calculated according to

$$S \propto |\sum_k \varphi_X(\mathbf{k})|^2. \quad (27)$$

The calculated total exciton oscillator strength drops smoothly with $V_g$ except near the half-filling $V_g$, where a dip appears because the exciton is screened by the free carriers [Fig. S15(c)]. The overall reduction of total oscillator strength with increasing $|V_g|$ is attributed to the influence of gate-induced vertical electric field, which increases linearly with $|V_g|$. The electric field can polarize the exciton wavefunction, decrease the electron-hole overlap, and hence reduce the oscillator strength. Such a reduction is approximately linear in $|V_g|$ for small $|V_g|$. We assume that the electric field modifies Eq. (27) to:

$$S \propto |\sum_k \varphi_X(\boldsymbol{k})|^2 (1 - 0.06|V_g|). \tag{28}$$

The modified results are shown in Fig. S15(c). The peak oscillator strength of the exciton is simply taken as $S/(\pi\Gamma)$, assuming that the exciton absorption peak has a Lorentzian line shape. As the FWHM ($\Gamma$) sharply increases at the half-filling $V_g$ positions, the peak oscillator strength shows a dip at the same positions. The simulation results for $S/(\pi\Gamma)$ are shown in Fig. S15(d).

For all the calculated results in Fig. S15(a-d), we have further broadened the curves by Gaussian convolution to account for the spatial inhomogeneity of the sample. The broadened curves (black lines) are used to compare with the experimental results in Fig. 3 of the main paper.

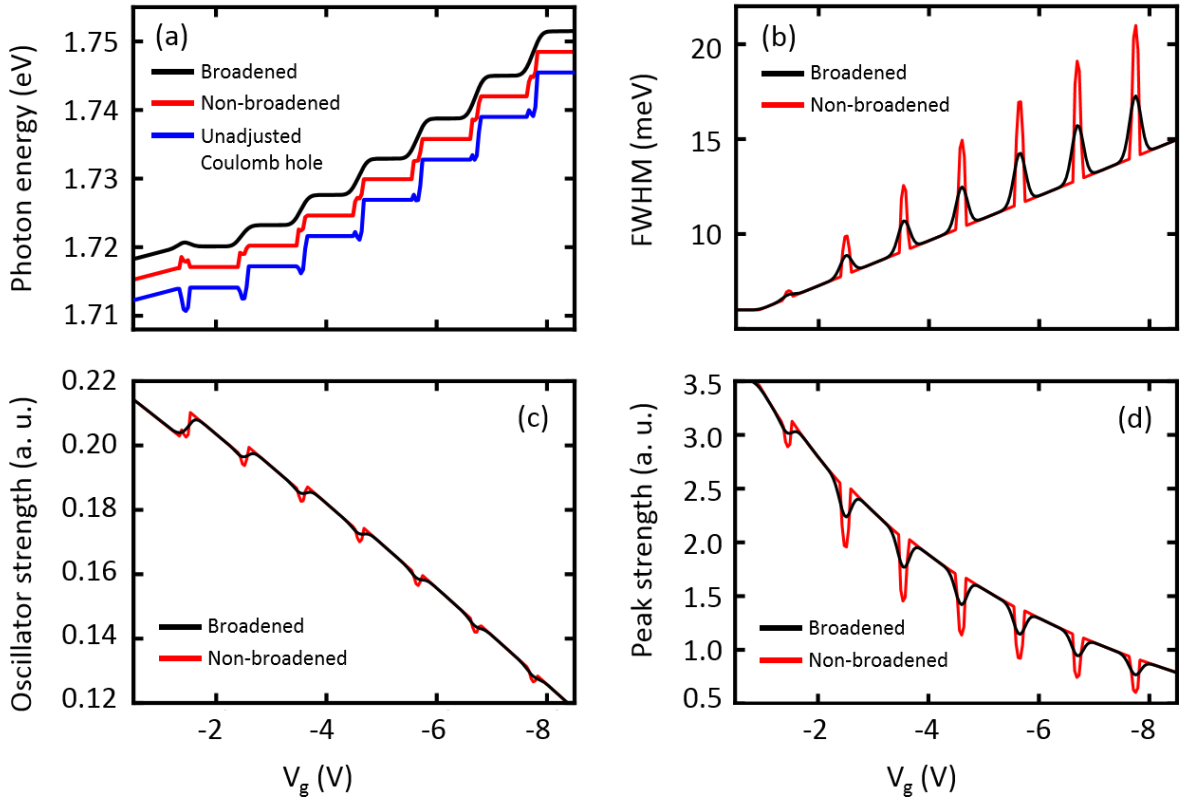

**Figure S15.** (a) Energy position, (b) full width at half maximum (FWHM), (c) total oscillator strength (S), and (d) peak oscillator strength of the $A^0$ exciton in monolayer WSe$_2$ at magnetic field B = 17.5 T versus the gate voltage ($V_g$) on the hole side. These results are obtained by the semi-empirical calculations described in the text. The non-broadened (red) and broadened (black) energy position in (a) are upshifted by 3 and 6 meV, respectively, for clarity. We have used the capacitance of our device to relate the gate voltage to the carrier density. The red curves are calculated for Landau levels with a uniform width of $\Gamma_{nn} = 0.002 R_x$. The black curves are the smoothened version of the red curves after an inhomogeneous broadening with a width of 0.1 V for $V_g$. The blue curve in (a) is calculated with the unadjusted reduction parameter η for the Coulomb-hole contribution. The broadened curves (black lines) are presented in Fig. 3 of the main paper to be compared with the experimental results.